\documentclass[%
 reprint,
superscriptaddress,
amsmath,amssymb,
pra,
]{revtex4-2}

\usepackage{amsmath,amssymb}
\usepackage{graphicx,float,color}
\usepackage{subcaption}
\usepackage{tikz}
\usetikzlibrary{positioning}
\usepackage[title]{appendix}
\usepackage{hyperref}
\usepackage{diagbox}
\def\bI{\mathbb{I}}
\def\bZ{\mathbb{Z}}
\def\cC{\mathcal{C}}
\def\cD{\mathcal{D}}
\def\cE{\mathcal{E}}

\def\iid{\mathsf{iid}}
\def\cN{\mathcal{N}}
\def\cO{\mathcal{O}}
\def\cP{\mathcal{P}}
\def\cS{\mathcal{S}}
\def\cR{\mathcal{R}}

\def\id{\mathsf{id}}
\def\dd{\diamondsuit}
\def\ol{\overline}
\def\ths{\thickspace}
\def\Tr{\mathsf{tr}}

\def\cL{\mathcal{L}}

\def\id{\mathcal{I}}

\def\frS{\mathfrak{S}}

\def\we{\mathsf{wt}}
\renewcommand{\Pr}{\mathsf{Pr}}
\def\wt{\widetilde}
\def\ot{\otimes}
\def\ts{\thinspace}
\def\cor{\mathsf{cor}}

\newcommand\ket[1]{|#1\rangle}
\newcommand\bra[1]{\langle#1|}
\def\cp{\widetilde{p}_{u}}

\newcommand{\figsdir}{figures/}

\definecolor{blue}{rgb}{0.5, 0, 0.13}

\begin{document}
\title{Efficient diagnostics for quantum error correction}

\author{Pavithran Iyer}
\affiliation{
Department of Applied Mathematics, University of Waterloo, Waterloo, Ontario N2L 3G1, Canada.
}
\affiliation{%
Institute for quantum computing, University of Waterloo, Waterloo, Ontario N2L 3G1, Canada.
}
\affiliation{%
Quantum Benchmark Inc., Kitchener, Ontario N2H 5G5, Canada.
}

\author{Aditya Jain}
\affiliation{
Department of Applied Mathematics, University of Waterloo, Waterloo, Ontario N2L 3G1, Canada.
}
\affiliation{%
Institute for quantum computing, University of Waterloo, Waterloo, Ontario N2L 3G1, Canada.
}
\affiliation{%
Quantum Benchmark Inc., Kitchener, Ontario N2H 5G5, Canada.
}
\affiliation{
Keysight Technologies Canada, Kanata, ON K2K 2W5, Canada.
}

\author{Stephen D. Bartlett}
\affiliation{%
Centre for Engineered Quantum Systems, School of Physics, University of Sydney, Sydney, New South Wales 2006, Australia.
}

\author{Joseph Emerson}
\affiliation{
Department of Applied Mathematics, University of Waterloo, Waterloo, Ontario N2L 3G1, Canada.
}
\affiliation{%
Institute for quantum computing, University of Waterloo, Waterloo, Ontario N2L 3G1, Canada.
}
\affiliation{%
Quantum Benchmark Inc., Kitchener, Ontario N2H 5G5, Canada.
}
\affiliation{
Keysight Technologies Canada, Kanata, ON K2K 2W5, Canada.
}

\begin{abstract}
Fault-tolerant quantum computing will require accurate estimates of the resource overhead, but standard metrics such as gate fidelity and diamond distance have been shown to be poor predictors of logical performance. We present a scalable experimental approach based on Pauli error reconstruction to predict the performance of concatenated codes. Numerical evidence demonstrates that our method significantly outperforms predictions based on standard error metrics for various error models, even with limited data. We illustrate how this method assists in the selection of error correction schemes.
\end{abstract}

\maketitle

Noise is pervasive in quantum processing, and must be overcome to achieve the disruptive capabilities of quantum computing. Fault tolerance guarantees reliable logical quantum computation in the presence of noise under prescribed conditions often oversimplified as achieving a threshold on gate error rates. However, achieving low logical error rates in practice is a significant challenge, in part because of the large overheads that are required in terms of the number of additional qubits and gates. The design and selection of an error correction strategy for a particular platform requires accurate prediction of its expected logical performance. For instance, in the presence of biased noise~\cite{AP08,RGBF17,TBF18,GM19,TBFB20,BATBFB21}, tailored codes have been shown to outperform traditional codes that are designed to correct unstructured noise. However, bias is only one of the exponentially many parameters that describe the noise on $n$ physical qubits. This work addresses the lack of tools for predicting the logical performance of a FT architecture based on a description of noise at the physical level.

The existing theoretical framework for choosing a FT scheme is centered around the fault tolerance accuracy threshold theorem \cite{AGP07,CTV17} which provides a threshold on the strength of physical noise below which reliable quantum computation can be guaranteed. However, directly applying the theorem to realistic noise has several challenges. One, the FT threshold is derived under oversimplified conditions that implicitly model a physical noise process as an incoherent error model with the same diamond distance. The leads to loose estimates of the logical performance when the noise has complex features such as coherence or strong correlations. Another is that diamond distance, which is usually invoked for assessing error rates in FT proofs, cannot be measured in a scalable way \cite{MC13}. It has been shown that the resource overheads for a FT architecture depend critically on the precise relationship between the architecture and the underlying error model. While there are several well-studied error metrics, none of them can accurately predict the logical error rate of a quantum code, with predictions varying by several orders of magnitude~\cite{IP17}. In this work we address this crucial deficiency prevalent in all of the known standard error metrics.

Here, we present a new figure of merit to predict the performance of concatenated codes, which can be measured efficiently using experimental protocols. As opposed to existing metrics such as average gate fidelity and diamond distance, our approach captures the interplay between the physical noise model and the choice of FT architecture. Our method leverages Randomized Compiling (RC)~\cite{WE16} to create an effective Pauli noise on the physical qubits, and then uses noise reconstruction (NR) techniques \cite{EWPM19,FW20,CER20} to estimate Pauli error probabilities. Using these data, we then design a \emph{logical estimator} that predicts the total probability of Pauli errors that a code cannot correct. While exactly computing this quantity is inefficient for a generic code, we introduce an efficient approximation to accurately estimate the total probability of uncorrectable errors for concatenated codes. We provide a bound on the efficiency and demonstrate the accuracy of our method through numerical simulations in several noise scenarios of interest. Finally, as an application, we demonstrate how the logical estimator pinpoints the selection of a suitable error correcting code for differing noise environments.

\section{Background} \label{sec:background}
A wide class of Markovian noise processes are formally described by completely positive trace preserving (CPTP) maps~\cite{C75}. There are several inequivalent ways to define the strength of noise modelled by a CPTP map $\cE$. Of these, the two most widely used to study fault tolerance are the \emph{average gate infidelity}~\cite{N96,B96,R01}: $
r(\cE) = 1 - \int d\psi~ \Tr(\ket{\psi}\bra{\psi} \ts \cE(\ket{\psi}\bra{\psi}))\ths,$
and the \emph{diamond distance}~\cite{K97,Kit97,Wat09,KSV02,G12}: $||\cE - \id||_{\dd} = \max_{\rho}||(\cE\ot \id)\rho - \rho||_{1} \ts$. While the average gate infidelity can be efficiently estimated using randomized benchmarking~\cite{GLN05,KLRB08,MGE11,MGE12}, the diamond distance satisfies mathematical properties that are needed in FT proofs~\cite{AB08,SDT07,AP09}. Other standard error metrics include the 2-norm~\cite{TB05}, Bures distance~\cite{B69}, Uhlmann fidelity~\cite{U76}, unitarity~\cite{WGHF15}, channel entropy~\cite{RZF11,Z14} and the adversarial error probability~\cite{AGP07}. None of these reflect the logical performance of a code \cite{IP17,I18}.

The net effect of a physical noise process $\cE_{0}$ together with a quantum error correction (QEC) routine using an $[[n,k]]$ stabilizer code~\cite{GotPhD97} is captured by the \emph{effective logical channel} $\cE^{s}_{1}$~\cite{RDM02} acting on an encoded state $\ol{\rho}$ as:
\begin{gather}
\cE^{s}_{1}(\ol{\rho}) = R_{s}\,\Pi_{s}\,\cE_{0}(\ol{\rho})\,\Pi_{s}\,R^{\dagger}_{s}/\Pr(s) \ths , \label{eq:effective_channel}
\end{gather}
where $\Pr(s)$ is the probability of measuring the syndrome outcome $s$, $\Pi_{s}$ is the syndrome projector and $R_{s}$ is the corresponding recovery. The average logical channel $\ol{\cE}_{1}$ is given by \cite{IP17,CWBL17}
\begin{gather}
\ol{\cE}_{1}(\ol{\rho}) = \sum_{s} \Pr(s) \cE^{s}_{1}(\ol{\rho})\ths . \label{eq:avgchan}
\end{gather}
While physical error rates are measured by noise-metrics on $\cE_{0}$, logical error rates are measured on $\ol{\cE}_{1}$.

Concatenated quantum codes are a popular family of codes of increasing sizes \cite{KL96}, and are often used to guarantee error suppression in fault tolerance proofs \cite{AGP07,JL14}. Physical qubits of a code $\cC_{\ell + 1}$ are encoded using a code $\cC_{\ell}$, for $1 \le \ell \le L{-}1$, yielding a \emph{level-$L$} concatenated code. The recursive encoding structure is represented by a tree where the $i$-th node at a depth $(L-\ell)$ denotes a quantum error correcting code block $\cC_{\ell, i}$. The sub-tree of the node is itself a concatenated code, denoted by $\cC^{\star}_{\ell, i}$, consisting of $(n^{\ell}{-}1)/(n{-}1)$ code blocks. There are $n{-}1$ independent stabilizer measurements corresponding to each of the code-blocks of $\cC^{\star}_{\ell,i}$. The resulting error syndrome $s(\cC^{\star}_{\ell,i})$ has $(n^{\ell}{-}1)$ bits, which can be grouped into subsets of $n{-}1$ bits that are identified by the code-blocks. We identify the subset of syndrome bits obtained by measurements on a code-block $\cC_{\ell,j}$ by $s(\cC_{\ell,j})$.

We consider the following iterative routine for quantum error correction in concatenated codes. For each level $\ell = 1,\ldots,L$: (i) syndromes are extracted for each code block $\cC_{\ell,1}, \ldots, \cC_{\ell, n}$, and (ii) a minimum-weight correction \cite{HL11} is applied in each case. Although we assume the popular choice of minimum-weight decoder in (ii), the methods prescribed in this work can be adapted to any lookup table decoder \cite{TS14}. Hence, the correction applied at any level depends on the syndrome history of the code blocks in the lower levels.

The effective channel for a level-$\ell$ concatenated code can also be computed in a recursive fashion, i.e., using eq.~(\ref{eq:effective_channel}) where $\cE_{0}$ is replaced by effective channel on the level-$(\ell-1)$ code blocks, i.e., $\cE^{s}_{\ell-1, 1} \ot \cdots \ot \cE^{s}_{\ell-1, n}$ \cite{RDM02,P06}. The performance of the level-$\ell$ concatenated code can be quantified~\cite{IP17} by the infidelity $
r(\ol{\cE}_{\ell}) = \sum_{s} \Pr(s) r(\cE^{s}_{\ell})$ of the average logical channel $\ol{\cE}_{\ell}$. For concatenated codes, it is possible to calculate both $\Pr(s)$ and the effective channel $\cE^{s}_{\ell}$. However, as the number of syndromes grow exponentially with the number of levels, Monte Carlo sampling techniques described in section \ref{sec:importance} of the appendix can be used to estimate this average.

\section{Methods} \label{sec:methods}
While the special setting of Pauli errors drastically simplifies the predictability problem, realistic noise processes are nonetheless poorly described by Pauli error models. To circumvent this problem, we recall a straightforward application of RC \cite{WE16} to fault tolerant circuits, that allows us to model the effect of complex noise processes by simple Pauli errors. In other words, RC ensures that there is no effect on the logical error rate from parameters of the physical channel other than the Pauli error probabilities. The physical twirling gates required to do RC can be absorbed into the logical gadgets of the FT circuits, at no additional cost in overhead. We provide explicit details of the procedure in the appendix section \ref{sec:qec_with_RC}.  

\begin{figure*}
\begin{tikzpicture}
\tikzstyle{every node}=[font=\scriptsize]
\node (a) at (-6,0) {\includegraphics[scale=0.40]{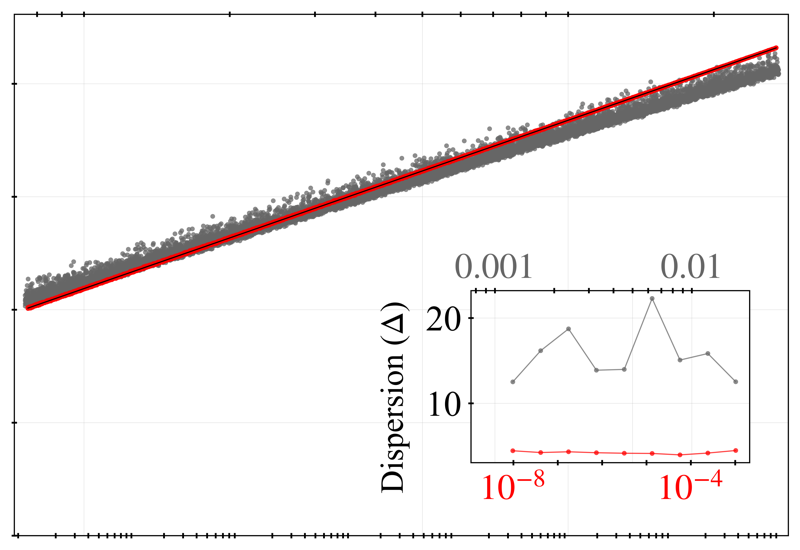}};
\node [above left = -0.85cm and -0.2cm of a] (ayt-3) {$3$};
\node [below = 0.38cm of ayt-3] (ayt-6) {$6$};
\node [below = 0.38cm of ayt-6] (ayt-9) {$9$};
\node [below = 0.38cm of ayt-9] (ayt-12) {$12$};
\node [below = 0.38cm of ayt-12] (ayt-15) {$15$};
\node [above left = -0.2cm and -1.2cm of a] (ait-3) {$0.001$};
\node [right = 1.3cm of ait-3] (ait-5_3) {$0.005$};
\node [right = 0.3cm of ait-5_3] (ait-2) {$0.01$};
\node [below left = -0.23cm and -1.2cm of a] (aut-8) {$10^{-8}$};
\node [right = 0.5cm of aut-8] (aut-6) {$10^{-6}$};
\node [right = 0.6cm of aut-6] (aut-4) {$10^{-4}$};
%
\node [right = 0.1cm of a] (b) {\includegraphics[scale=0.40]{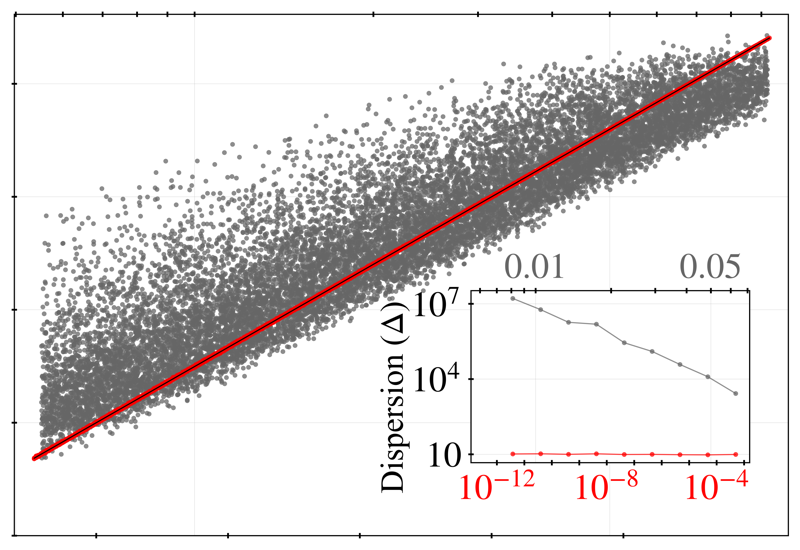}};
\node [above left = -0.85cm and -0.2cm of b] (byt-3) {$3$};
\node [below = 0.38cm of byt-3] (byt-6) {$6$};
\node [below = 0.38cm of byt-6] (byt-9) {$9$};
\node [below = 0.38cm of byt-9] (byt-12) {$12$};
\node [below = 0.38cm of byt-12] (byt-15) {$15$};
\node [above left = -0.2cm and -0.8cm of b] (bit-5_3) {$0.005$};
\node [right = 0.35cm of bit-5_3] (bit-2) {$0.01$};
\node [right = 2.07cm of bit-2] (bit-5_2) {$0.05$};
\node [below left = -0.25cm and -1.45cm of b] (but-12) {$10^{-12}$};
\node [right = 0.85cm of but-12] (but-8) {$10^{-8}$};
\node [right = 0.9cm of but-8] (but-4) {$10^{-4}$};
%
\node [right = 0.1cm of b] (c) {\includegraphics[scale=0.40]{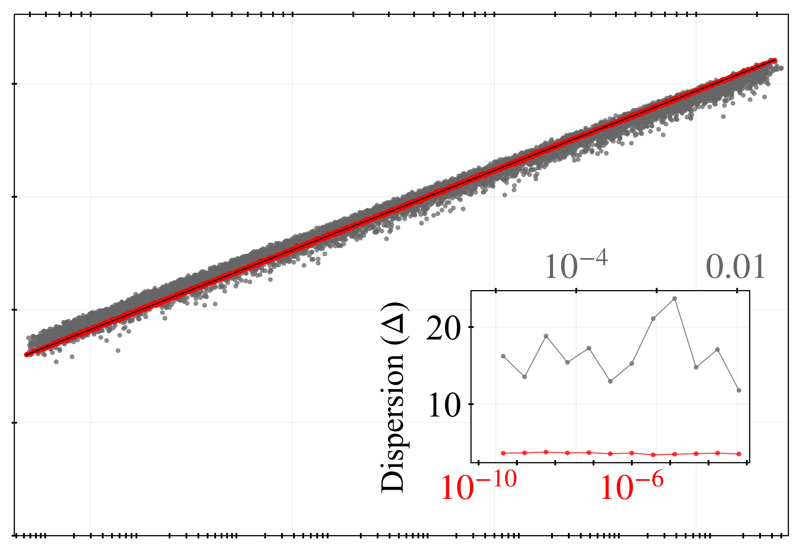}};
\node [above left = -0.85cm and -0.2cm of c] (cyt-3) {$3$};
\node [below = 0.38cm of cyt-3] (cyt-6) {$6$};
\node [below = 0.38cm of cyt-6] (cyt-9) {$9$};
\node [below = 0.38cm of cyt-9] (cyt-12) {$12$};
\node [below = 0.38cm of cyt-12] (cyt-15) {$15$};
\node [above left = -0.2cm and -1.3cm of c] (cit-5) {$10^{-5}$};
\node [right = 0.55cm of cit-5] (cit-4) {$10^{-4}$};
\node [right = 0.47cm of cit-4] (cit-3) {$0.001$};
\node [right = 0.52cm of cit-3] (cit-2) {$0.01$};
\node [below left = -0.25cm and -1.3cm of c] (cut-10) {$10^{-10}$};
\node [right = 0.37cm of cut-10] (cut-8) {$10^{-8}$};
\node [right = 0.45cm of cut-8] (cut-6) {$10^{-6}$};
\node [right = 0.45cm of cut-6] (cut-4) {$10^{-4}$};
%
\node [above left = 0.15cm and -5.2cm of a] (atopx) {Physical avg.~gate infidelity w/o RC};
\node [below = 0.2cm of a] (abottomx) {Logical estimator with RC};
\node [rotate=90, above left = -0.3 cm and 0.15cm of a] (aylabel) {$-\log_{10}(\text{Logical infidelity})$};
\node [below = -0.1cm of abottomx] (acap) {(a)};
\node [above left = 0.15cm and -5.2cm of b] (btopx) {Physical diamond distance w/o RC};
\node [below = 0.2cm of b] (bbottomx) {Logical estimator with RC};
\node [below = -0.1cm of bbottomx] (bcap) {(b)};
\node [above = 0.15cm of c] (ctopx) {Physical avg.~gate infidelity};
\node [below = 0.2cm of c] (cbottomx) {Logical estimator};
\node [below = -0.1cm of cbottomx] (ccap) {(c)};
\end{tikzpicture}

\vspace{-0.2cm}

\caption[predictability]{Predictability of logical infidelity for level-2 concatenated Steane code. Figures 1(a,b) compare the predictive powers of our logical estimator (red) against two standard error metrics (gray): the average gate infidelity (a) and the diamond distance (b), under an ensemble of 18000 CPTP maps. Each point $p = (x_{p}, y_{p})$ corresponds to a physical noise process; $x_{p}$ is its physical error metric and $y_{p}$, its logical error rate. The dispersion of points, quantified as $\Delta$ in the insets, indicates the predictive power of the physical error metric. While logical error rates can vary over several orders of magnitude with respect to standard error metrics, our logical estimator is strongly correlated with the logical error rate. (c): Correlated Pauli error model.}
\label{fig:pred}
\end{figure*}
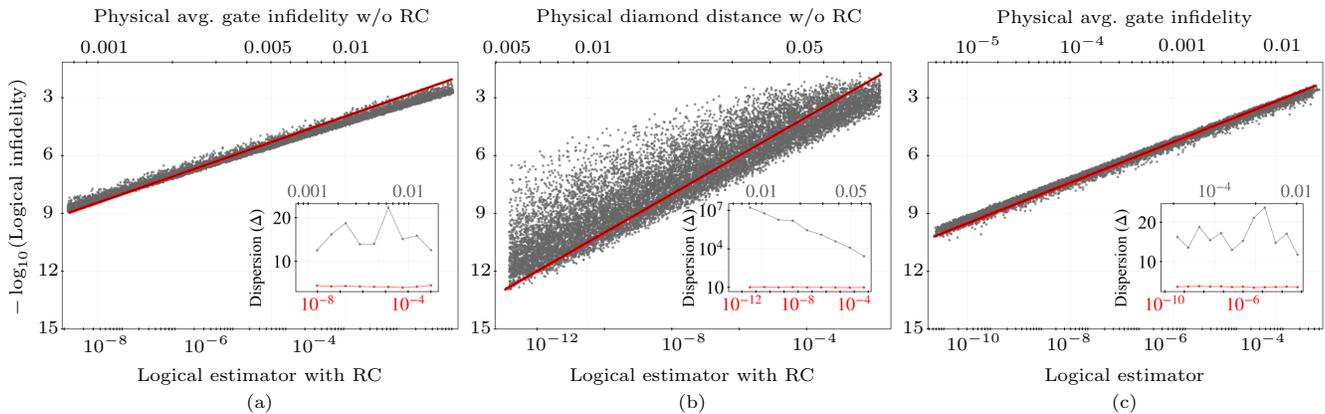

With a noise model described by Pauli errors, we first develop the background needed to define notion of a \emph{logical estimator} that can accurately predict the logical error rate. A stabilizer code and a decoder pair is designed to correct a target set of errors $\cE_{C}$, called \emph{correctable errors}~\cite{S06,R12}. Errors not in this set, i.e., uncorrectable errors, contribute to the logical error rate. Ideally, we want to estimate the total probability of all uncorrectable errors, which can be obtained by adding the probabilities of correctable errors and subtracting the result from one:
\begin{gather}
p_{u}(\cC) = 1 - \sum_{E \in \cE_{C}}\chi_{E, E} \ths ,\label{eq:pu}
\end{gather}
where $\chi_{P, Q}$ is the element corresponding to the Pauli operators $P$ and $Q$, in the chi-matrix representation \cite{WBC15} of the physical noise process. For Pauli error models, $p_{u}$ is identical to the logical infidelity, while for generic CPTP maps, a precise relationship is presented in section \ref{sec:loginfid} of the appendix. While on the one hand, the average gate infidelity $r$ accounts for the effect of only the trivial correctable error $\bI$, $p_{u}$ on the other hand captures all the degrees of freedom that are relevant to the logical error rate. However, an exact computation of $p_{u}$ is intractable in general.

A correctable error $E_{\ell}$ for the concatenated code $\cC^{\star}_{\ell}$ falls into one of the two categories: either (i) it is corrected within the lower level code-blocks $\cC^{\star}_{\ell-1, 1}, \ldots, \cC^{\star}_{\ell-1, n}$, or (ii) it has a non-trivial correction applied by the decoder of the level$-\ell$ code-block $\cC_{\ell,1}$. Adding up the contributions to $1 - p_{u}(\cC^{\star}_{\ell})$ from cases (i) and (ii) respectively, we find
\begin{gather}
1 - p_{u}(\cC^{\star}_{\ell,1}) = \prod_{j=1}^{n} (1-p_{u}(\cC^{\star}_{\ell-1,j})) + \Gamma(\cC^{\star}_{\ell,1}) \ths , \label{eq:pu}
\end{gather}
where $\Gamma(\cC^{\star}_{\ell,1}) = \sum_{E \ts\in\ts\cE_{\cC}\backslash \bI} \Pr(\otimes_{j=1}^{n} E_{\ell-1,j})$. An exact computation of $\Gamma(\cC^{\star}_{\ell,1})$ involves enumerating all possible syndrome outcomes for the level$-\ell$ concatenated code.

Our \emph{logical estimator}, denoted by $\cp$, is the result of estimating $p_{u}$ using an efficient approximation for $\Gamma(\cC^{\star}_{\ell,1})$ for concatenated codes. In particular, we use a coarse grained estimate of the probability of a syndrome outcome -- a joint probability distribution over $\cO(n^{\ell})$ syndrome bits -- calculated as a product of marginal probability distributions over the $n$ code blocks at level $(\ell-1)$. This procedure is recursed through the $\ell$ levels of the concatenated code. A detailed derivation of the logical estimator is provided in section \ref{sec:uncorr} of the appendix.

For i.i.d Pauli error models with sufficiently small single-qubit infidelity $r_{0}$, the quality of approximation is: $|\ol{r}_{\ell} - \cp| \leq n_{C}^{\ell+1} r_{0}^{2+\lfloor(d_{C}+1)/2\rfloor}$. Here, $d_{C}$ and $n_{C}$ describe the distance and the size of a codeblock of a level$-\ell$ concatenated code. For instance, using an i.i.d depolarizing error model with $r_{0} = 10^{-3}$ and the level-$2$ concatenated Steane code, the above expression yields $|\ol{r}_{2} - \cp| \leq 5 \times 10^{-10}$. This is validated by numerics: $\cp = 4.24 \times 10^{-9}$ and $\ol{r}_{2} = 4.20 \times 10^{-9}$. Notably, the time complexity of computing $\cp$ for the concatenated code:  $\cO(4^{n_{C}+\ell} \ths n^{\ell})$, scales polynomially in the total number of physical qubits, i.e., scaling as $n^{\ell}$, whereas an exact computation of $p_{u}$ would scale doubly exponentially in $\ell$. An analysis of the quality and efficiency of the approximation, can be found in sections \ref{sec:pu_quality} and \ref{sec:pu_time_app}, respectively, of the appendix.

\section{Results and discussion} \label{sec:numerics}
We provide numerical evidence to highlight the improvements offered by our methods developed for optimizing FT schemes. We begin with the task of accurately predicting the performance of concatenated Steane codes. We perform numerical simulations of quantum error correction in the RC and non-RC settings under a large ensemble of random CPTP maps applied to the physical qubits. Following Ref.~\cite{IP17}, we generate a single qubit CPTP map $\cE$ from its Stinespring dilation: a random unitary matrix $U$ of size $(8\times 8)$, given by $U = e^{-iHt}$ for a complex Hermitian matrix $H$ whose entries are sampled from a Gaussian distribution of unit variance, centred at 0. We vary the time parameter $t$ between $0.001$ and $0.1$ to vary the noise strength.

Figure \ref{fig:pred} shows that logical error rates can vary wildly across physical noise processes with fixed infidelity and diamond distance respectively, in agreement with the work in Ref.~\cite{IP17}. The variation, captured by the amount of dispersion in the scatter plots, is quantified using a simple measure -- the ratio of the minimum and the maximum logical error rates across channels of similar physical error rate, denoted by $\Delta$. In other words, we partition the range of physical error rates into bins $b_{i}$ and use $\Delta(b_{i})$ to quantify the amount of dispersion: $\Delta(b_{i}) = (1/|b_{i}|) \ths (\max_{p \in b_{i}}y_{p})/(\min_{p \in b_{i}}y_{p})$, where $|b_{i}|$ is the number of channels in the bin $b_{i}$. The large fluctuations in the logical error rates can be attributed to two extreme features of the error-metrics. While infidelity controls only one parameter out of the many that specify a noise process, diamond distance, on the other hand, suffers from being sensitive to the details of a noise process that are irrelevant to the logical error rate. In addition, standard error metrics can only reveal intrinsic properties of the underlying noise process, that are agnostic to the choice of an error correcting code.

Logical estimator with RC, in contrast, is very highly correlated with the logical error rate. This improvement can be attributed to two features. First, RC provides a drastic reduction from $\cO(12^{n})$ parameters that specify an $n-$qubit Markovian noise process to $\cO(4^{n})$ Pauli error probabilities. Second, unlike standard error metrics, $\cp$ carefully accounts for Pauli error probabilities that contribute to the logical error rate. Section \ref{sec:results_coherent} of the appendix shows drastic gains in predictability using the logical estimator with RC, over standard error metrics, for the class of coherent errors.

The special setting of i.i.d noise hides the drastic advantages provided by $\cp$ in predicting logical infidelity because the dominant contribution to $\cp$ comes from $\chi_{0,0}$, which is also well captured by $r$. However, for correlated error-models, given only $\chi_{0,0}$, the uncertainly on the logical error rate ranges between the extremities, 0 and 1, achieved when all the multi-qubit errors are correctable, and uncorrectable, respectively. While $r$ is completely insensitive to either of these scenarios, $\cp$ in contrast helps distinguish between them, thereby providing a far more accurate estimate of the logical error rate.

We support the above argument by numerical studies of correlated Pauli error models generated from a convex combination of an i.i.d process of infidelity $r_{0}$ and  multi-qubit interactions. While the i.i.d component $\cE_{\iid}$ is specified by single qubit error probabilities, multi-qubit interactions are specified by an arbitrary subset $\frS$, so, $\cE_{\cor}(\rho) = \sum_{P\in\frS}\chi_{P,P}P\,\rho\,P$, where $\chi_{P,P}$ is sampled from the normal distribution with mean and variance $4^{n} r_{0}$. The combined Pauli error model is therefore given by $\cE(\rho) = q \cE_{\iid}(\rho) + (1 - q) \cE_{\cor}(\rho)$, where $0\leq q\leq 1$. Explicitly setting $\chi_{0,0}$ followed by appropriate normalization, ensures that the infidelity of the above noise model is $r_{0}$. Figure \ref{fig:pred}(c) highlights the importance of the logical estimator over the standard infidelity error metric for predicting the performance of the concatenated Steane code under correlated Pauli noise processes.

\subsection{Logical estimator using limited NR data} \label{sec:partial_NR}
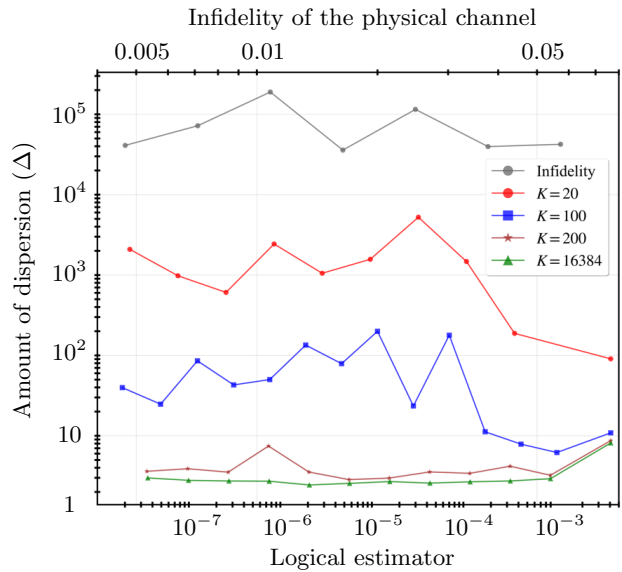
\begin{figure}
\begin{tikzpicture}
\node (limitedNR) at (-4.8,0) {\includegraphics[scale=0.55]{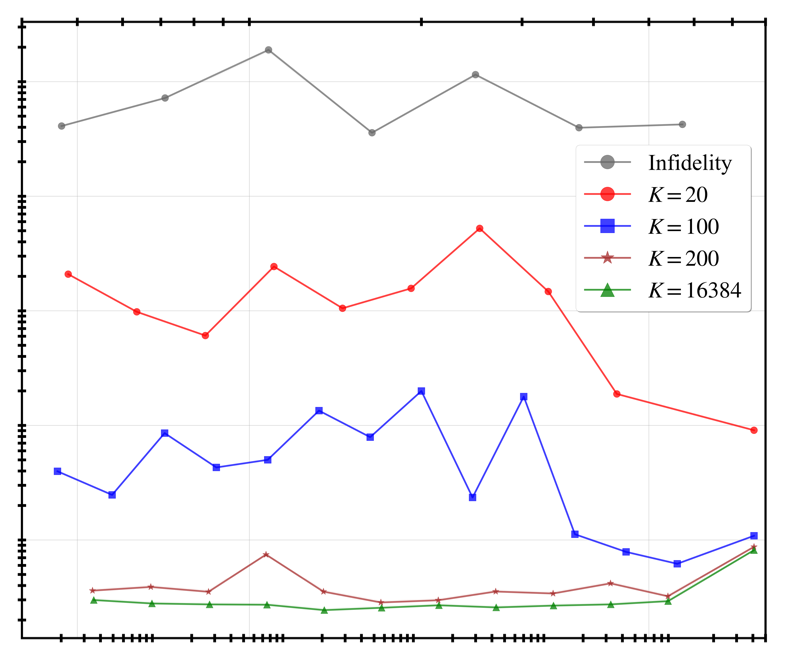}};
\node [above left = -1.1cm and -0.3cm of limitedNR] (LDyt5) {$10^{5}$};
\node [below = 0.55cm of LDyt5] (LDyt4) {$10^{4}$};
\node [below = 0.55cm of LDyt4] (LDyt3) {$10^{3}$};
\node [below = 0.55cm of LDyt3] (LDyt2) {$10^{2}$};
\node [below = 0.65cm of LDyt2] (LDyt1) {$10$};
\node [below = 0.5cm of LDyt1] (LDyt0) {$1$};
\node [above left = -0.25cm and -1.4cm of limitedNR] (LDit-5_3) {$0.005$};
\node [right = 0.72cm of LDit-5_3] (LDit-2) {$0.01$};
\node [right = 2.91cm of LDit-2] (LDit-5_2) {$0.05$};
\node [below left = -0.3cm and -2.1cm of limitedNR] (LDut-7) {$10^{-7}$};
\node [right = 0.3cm of LDut-7] (LDut-6) {$10^{-6}$};
\node [right = 0.3cm of LDut-6] (LDut-5) {$10^{-5}$};
\node [right = 0.3cm of LDut-5] (LDut-4) {$10^{-4}$};
\node [right = 0.3cm of LDut-4] (LDut-3) {$10^{-3}$};
\node [above = 0.15cm of limitedNR] (topx) {Infidelity of the physical channel};
\node [below = 0.15cm of limitedNR] (bottomx) {Logical estimator};
\node [rotate=90, above left = 0.1 cm and 0.2cm of a] (ylabel) {Amount of dispersion $(\Delta)$};
\end{tikzpicture}

\vspace{-0.2cm}

\caption[infid vs pu]{Accuracy of the logical estimator based on limited NR data, using a level-2 concatenated Steane code for an ensemble of about 15000 random correlated Pauli channels. The accuracy, quantified by $\Delta$, improves sharply with the number of Pauli error rates ($K$) extracted using NR. We observe that for $K = 200$, which is about $1.2\%$ of all Pauli error rates on the Steane code block, the accuracy closely matches the logical estimator computed using all NR data, i.e., $K = 4^{7}$.}
\label{fig:partial_NR}
\end{figure}

Even in the absence of correlations across the $n-$qubit code blocks of a concatenated code, we require $\cO(4^{n})$ Pauli error rates from NR to compute $\cp$. Extracting this exponential sized NR dataset is a challenge for experimentalists. Refs.~\cite{HYF20,CER20} describe how to extract the leading $K$ Pauli error probabilities in a noise process, where $K \ll 4^{n}$. We want to combine a handful of leading Pauli error rates extracted by NR with a simple method to extrapolate the remaining ones. We define the probability of a Pauli error $Q$ that is not given in the NR dataset as
\begin{gather}
\Pr(Q) = (1-r_{0})^{n - \we(Q)}\left(r_{0}/3\right)^{\we(Q)} \ths , \label{eq:limited_heuristic}
\end{gather}
where $\we(Q)$ is the Hamming weight of $Q$, and $r_{0}$ is derived from the infidelity of the noise process: $r= 1 - (1 - r_{0})^{n}$. We construct an adversarial example of an error model where the above extrapolation is unlikely to perform well by setting some multi-qubit error probabilities that violate eq.~(\ref{eq:limited_heuristic}). Furthermore, when errors are sampled uniformly from the set of correctable and uncorrectable errors, we observe maximum fluctuations in the logical error rate. However, Fig.~\ref{fig:partial_NR} presents strong numerical evidence indicating that the simple extrapolation works well in practice even for the adversarial example.

\subsection{Code selection} \label{sec:code_selection}
Selecting a quantum error correcting code that has the smallest logical error rate under an existing physical noise process is a crucial step in optimizing resources for fault tolerance. To demonstrate the efficacy of the logical estimator for this problem, we consider an example of an error model and two different error-correcting codes: (i) concatenated Steane code and (ii) concatenated version of a $[[7,1,3]]$ code used in Ref.~\cite{RGBF17} that we refer to as a \emph{Cyclic code}. The error model is obtained from a Pauli twirl on the i.i.d application of the CPTP map $\cE: \rho \mapsto p_{I}\rho + \sum_{Q \in \{X, Y, Z\}}p_{Q} e^{-i \theta Q} \rho e^{i \theta Q}$, where $p_{Y} = p_{X}p_{Z}, p_{I} = 1 - p_{X} - p_{Y} - p_{Z}$ and set a bias specified by $\eta = p_{Z}/p_{X}$. Based on Ref.~\cite{RGBF17}, we expect the Steane code to outperform the Cyclic code in one noise regime, and the converse in a different regime. Our tool is successful if it produces a lower value of $\cp$ for the code with lower logical infidelity, for any noise rate. Lastly, to compute the logical estimator as well as the logical error rate estimates, we use a bias-adapted minimum-weight decoder that assigns weights $\eta$, $\eta$, and $1$ to each Pauli error of type $X$, $Y$, and $Z$, respectively.

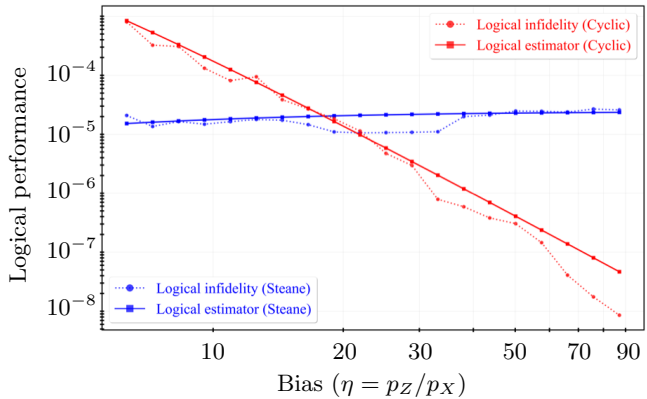
\begin{figure}
\begin{tikzpicture}
\node (codeselection) at (-4.4,0) {\includegraphics[scale=0.55]{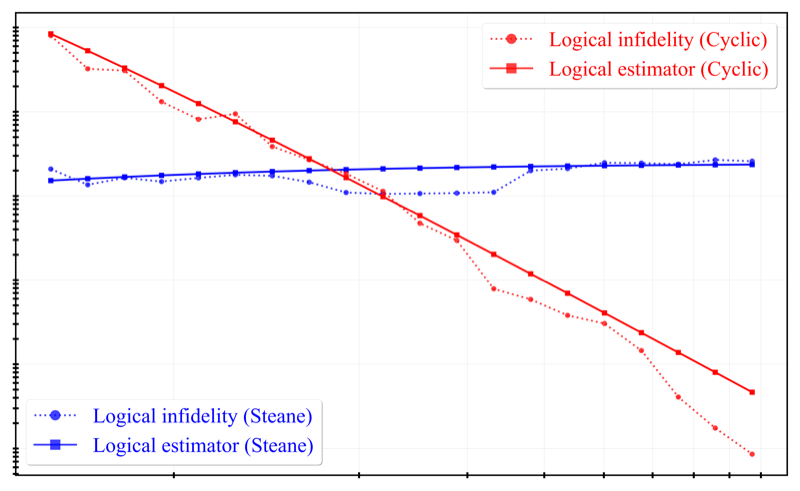}};
\node [above left = -1.35cm and -0.3cm of codeselection] (CS-4) {$10^{-4}$};
\node [below = 0.3cm of CS-4] (CS-5) {$10^{-5}$};
\node [below = 0.25cm of CS-5] (CS-6) {$10^{-6}$};
\node [below = 0.3cm of CS-6] (CS-7) {$10^{-7}$};
\node [below = 0.25cm of CS-7] (CS-8) {$10^{-8}$};
\node [below left = -0.2 and -2cm of codeselection] (B10) {$10$};
\node [right = 1.2cm of B10] (B20) {$20$};
\node [right = 0.45cm of B20] (B30) {$30$};
\node [right = 0.7cm of B30] (B50) {$50$};
\node [right = 0.3cm of B50] (B70) {$70$};
\node [right = 0.1cm of B70] (B90) {$90$};
\node [below = 0.2cm of codeselection] (bottomx) {Bias $(\eta = p_{Z}/p_{X})$};
\node [rotate=90, above left = -0.5 cm and 0cm of a] (ylabel) {Logical performance};
\end{tikzpicture}

\vspace{-0.2cm}

\caption{Using the logical estimator to select an optimal code. The above figure demonstrates the use of our tool in selecting an optimal error correcting code under a biased Pauli error model. The choices of codes is between the level$-3$ concatenated versions of the Steane code and the cyclic code. While the solid lines depict the values of the logical estimator $\cp$, the dashed lines correspond to the logical error rates estimated using numerical simulations. We observe that $\cp$ accurately selects the optimal code for all noise rates.}
\label{fig:code_selection}
\end{figure}

Our results presented in Fig.~\ref{fig:code_selection} show that the logical estimator correctly identifies the optimal code for all values of the physical error rate (bias). Furthermore, it also replicates the functional form of the logical error rate, showing that the performance gain from the Cyclic code over the Steane code increases with the bias.

\section{Conclusions} \label{sec:conclusion}
We have shown how experimental data from NR, even limited data, can be used to successfully predict the logical performance of FT architectures based on concatenated codes.  It can be used to precisely and efficiently estimate the resource overhead required to achieve a target logical error rate \cite{JROC17,RGBF17,NDDV19} for implementing quantum algorithms. Along with informing the choice of an optimal code for an underlying physical noise process, the logical estimator provides directives for other components in a FT scheme, such as a decoder. Different lookup table decoders can be compared using our logical estimator, similar to the work in Refs.~\cite{CR18,SJG20,DPMCS20}.

Our scheme relies on RC to yield a Pauli error model, and although in theory this requires twirling with the full Pauli group, it has been observed in practice that a handful of random compilations of the original circuit are sufficient to achieve an almost Pauli-like effective noise process \cite{HNMVM20,SLSKN21}. A natural question that follows is whether RC also mitigates the impact of physical noise on the logical qubit. There is no persistent trend across the general class of Markovian noise processes, and in some cases, RC degrades the performance of the code. Developing noise tailoring techniques that guarantee an improvement to the performance as well as predictability is an interesting problem for future research.

Although the methods and techniques presented in the paper address generic noise processes, there are a number of roadblocks in broadening the scope of this study beyond concatenated codes, where the complexity of computing the logical estimator grows exponentially with the size of the code. In addition, further research is needed to extend these ideas to the context of multiple logical qubits.

\begin{acknowledgements}
We thank Daniel Gottesman, Joel Wallman and Stefanie Beale for their valuable inputs. This research was undertaken thanks in part to funding from the Canada First Research Excellence Fund. Research was partially sponsored by the ARO and was accomplished under Grant Number: W911NF-21-1-0007.  SDB acknowledges support from the Australian Research Council (ARC) via the Centre of Excellence in Engineered Quantum Systems (EQuS) project number CE170100009.  
\end{acknowledgements}

\begin{appendix}
\begin{widetext}

\section{Logical fidelity and correctable errors} \label{sec:loginfid}
The average logical channel $\cE^{s}_{1}$, defined in eq. \ref{eq:effective_channel}, summarizes the effect of quantum error correction on a physical noise process $\cE_{0}$ affecting an encoded state $\ol{\rho}$. In this section, we derive a closed form expression for the average logical channel in terms of the physical channel and the error correcting code parameters. Similar derivations have appeared in \cite{CWBL17,GB15,RDM02}, however, we present ours for the sake of completeness.

The action of the average logical channel defined in eq. \ref{eq:avgchan} on the logical state is
\begin{flalign}
\ol{\cE}_{1}(\ol{\rho}) = &\sum_{s}\Pr(s)\cE^{s}_{1}(\ol{\rho}) \nonumber , \\
= &\sum_{s} R_{s}\Pi_{s} \cE_{0}(\ol{\rho}) \Pi_{s}R_{s} \nonumber\\
= &\sum_{s} \sum_{i,j} \chi_{i,j} R_{s}\Pi_{s} P_{i} \ol{\rho} P_{j} \Pi_{s}R_{s} \nonumber \\
= &\sum_{s}\sum_{\substack{i,j \\ s(P_{i}) = s(P_{j}) := s}}\chi_{i,j} R_{s} P_{i} \ol{\rho} P_{j} R_{s} \ths , \label{eq:average_eff_state_chi_no_projs}
\end{flalign}
where in the last line we used the fact that $\Pi_{s} P_{i} = P_{i} \Pi_{s \oplus s(P_{i})}$. In other words, whenever $s \neq s(P_{i})$, the corresponding projector $\Pi_{s \oplus s(P_{i})}$ annihilates the encoded state $\ol{\rho}$.

The chi matrix $\ol{\chi}$ of the effective logical channel defined by
\begin{gather}
\cE_{1}(\ol{\rho}) = \sum_{l,m} \ol{\chi}_{lm} \ol{P}_{l}\ol{\rho}\ol{P}_{m} , \label{eq:logical_chi_channel}
\end{gather}
where $\ol{P}_{l}$ and $\ol{P}_{m}$ are logical operators of the code; can be extracted from eq. \ref{eq:average_eff_state_chi_no_projs}.

The total probability of errors successfully corrected by the decoder: $\ol{\chi}_{00}$, can be estimated from the following observation. An error whose syndrome is $s$ is corrected if the net effect of applying the error along with a recovery prescribed by the decoder results in an effective action of a stabilizer. In other words, all the terms in eq. \ref{eq:average_eff_state_chi_no_projs} where $R_{s}P_{i}$ and $P_{j}R_{s}$ are stabilizers contribute to $\ol{\chi}_{00}$. So,
\begin{gather}
\ol{\chi}_{0,0} = \sum_{\substack{E,E^{\prime} \in \cE_{C} \\ s(E) = s(E^{\prime}) \ts , \ts \ol{E} = \ol{E}^{\prime}}} \phi(E) \ths  \phi^{\star}(E^\prime) \ths \chi_{E,E^{\prime}}  \ths ,\label{eq:chi00_log_phys}
\end{gather}
where $\ol{E}$ is the logical component in the decomposition of $E$ with respect to the Stabilizer group and $\phi(E)$ is specified by $R_{s(E)} E = \phi(E) \ths S$ for any Pauli error $E$ and some stabilizer $S$. The average logical infidelity $\ol{r}$ is then given by $1 - \ol{\chi}_{00}$.

When a Pauli error is not correctable, the effect of applying a recovery yields a logical operator. Hence, in general
\begin{gather}
\ol{\chi}_{l,m} = \sum_{\substack{E,E^{\prime} \in \cE_{C} \\ s(E) = s(E^{\prime}) \ts , \ts \ol{E} = \ol{E}^{\prime}}} \phi(E, l) \ths  \phi^{\star}(E^\prime, m) \ths \chi_{E \ol{P}_{l},\ol{P}_{m}E^{\prime}} \ths .\label{eq:chi_lm_log_phys}
\end{gather}
where $R_{s(E)} \ths E \ths \ol{P}_{l} = \phi(E,s) \ths S$, for $l \in \{0,1,2,3\}$, any Pauli error $E$ and some stabilizer $S$.

\section{Quantum error correction with randomized compiling} \label{sec:qec_with_RC}
In this section, we discuss how randomized compiling (RC) can be performed in fault tolerant circuits. Note that a Pauli error $P$ can be decomposed with reference to a stabilizer code: $P = \ol{P}~S_{P}~E_{P}$, where $S_{P}$ is an element of the stabilizer group $\cS$, $\ol{P}$ is a logical Pauli operator in $\cL = \cN(\cS)/\cS$, and $E_{P}$ is an element of $\cN(\cL)/\cS$, usually called a pure error \cite{LB13,P06}. Unlike pure errors, stabilizers and logical operators commute with QEC routines. A Pauli error $P$ can be compiled into QEC resulting in a new quantum error correction routine $QEC(P)$ in which the input to the decoder corresponding to a syndrome outcome $s$ is $s \oplus s(P)$ \cite{DA07,CIP18}.

In fault tolerant circuits, each logical gate $\ol{G}$ is sandwiched between QEC routines. Following the prescription in \cite{WE16}, we divide logical gates into two sets: $\cS_{1}$ and $\cS_{2}$, calling them easy and hard gates respectively. A crucial requirement for $\cS_{1}$ and $\cS_{2}$ is
\begin{gather}
\ol{G}~T~\ol{G}^{\dagger}~QEC = QEC(T)~\ol{C} \ths \label{eq:P_map_easy}
\end{gather}
for all easy logical gates $\ol{C}\in\cS_{1}$, $n-$qubit Pauli gates $T$ and hard gates $\ol{G}$. Recall that $QEC(T)$ refers to the compilation of the Pauli gate $T$ in the QEC routine, discussed in sec. \ref{sec:background}. It follows from
\begin{flalign}
\ol{G}~T~\ol{G}^{\dagger}~QEC &= \ol{G}~\ol{T}~\ol{G}^{\dag}~S_{T}~E_{T}~QEC \label{eq:apply_tls} \\
&= \ol{G}~\ol{T}~\ol{G}^{\dag}~QEC(E_{T}) \ths , \label{eq:easy_gadget_compiled}
\end{flalign}
where, in eq. \ref{eq:apply_tls} we have used the decomposition of Pauli gates with reference to a stabilizer code. Note that the expression $\ol{G} ~\ol{T}~\ol{G}^{\dag}$ in eq. \ref{eq:easy_gadget_compiled} is guaranteed to be an easy gate for a choice of easy and hard gate sets in \cite{WE16}.

Fig.~A.2(a) shows a canonical presentation of a quantum circuit, where the $k$-th clock cycle is composed of an easy gate $\ol{C}_{k}$ and a hard gate $\ol{G}_{k}$, sandwiched between QEC routines. Noise processes affecting easy and hard gates are denoted by $\cE_{1,k}$ and $\cE_{2,k}$ respectively. These complex processes can be tailored to Pauli errors by inserting Pauli gates $T_{1,k}, T^{\dag}_{1,k},T_{2,k}, T^{\dag}_{2,k}$. However, to guarantee that they be applied in a noiseless fashion, we compile them into the existing gates in the fault tolerant circuit. This is achieved in two steps. First, $T^{\dagger}_{1,k}$ and $T_{2,k}$ are compiled into QEC following $\cE_{1,k}$, resulting in $QEC(T^{\dagger}_{1,k}T_{2,k})$. Second, $T^{\dagger}_{2,k}$ is propagated across $\ol{G}_{k}$, and compiled with $QEC~\ol{C}_{k+1}T_{k+1}$, resulting in a \emph{dressed gate} $\ol{C^{D}_{k+1}} = \ol{G}_{k}~T_{k}~\ol{G}_{k}^{\dagger} \ths QEC \ths \ol{C}_{k+1} \ths T_{k+1}.$ It follows from eq. \ref{eq:P_map_easy} that $\ol{C^{D}_{k+1}}$ is equivalent to quantum error correction followed by an easy gate.

Fig.~A.2(c) shows the result of compiling all of the twirling gates into the easy gates and quantum error correction routines. Note that the compiled circuit is logically equivalent to the original circuit in the absence of noise. However, in the presence of noise, the average output of the circuit is dictated by the performance of $QEC(T)$ averaged over the different choices of Pauli gates $T$. This is what we refer to as QEC in the \emph{RC setting}. In practice, this average performance can be achieved by repeating every iteration (shot) of the algorithm with a different Pauli operation compiled into the constituent QEC routines. For the purpose of numerical simulations in this paper, we have used the performance of the QEC routine under the twirled noise process, as a proxy to the performance of QEC in the RC setting.

\begin{figure*}
\begin{tikzpicture}
\node (a) at (-2,0) {\includegraphics[scale=0.1]{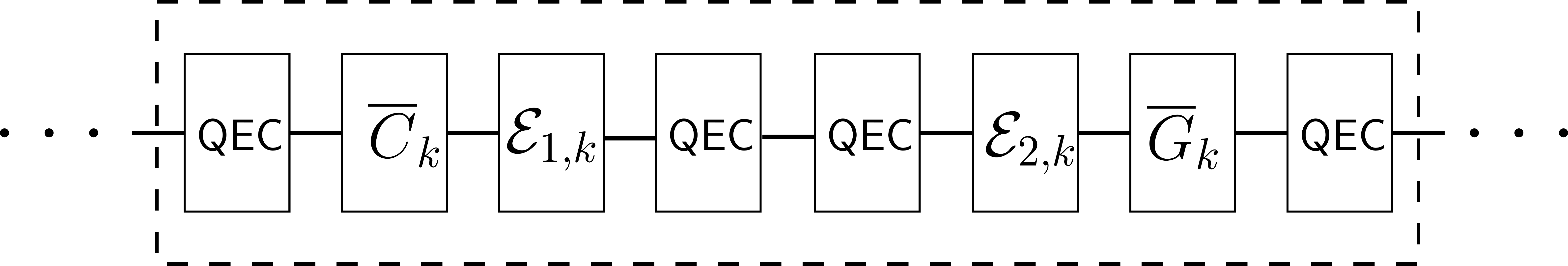}};
\node [below = 0.15cm of a] (b) {\includegraphics[scale=0.1]{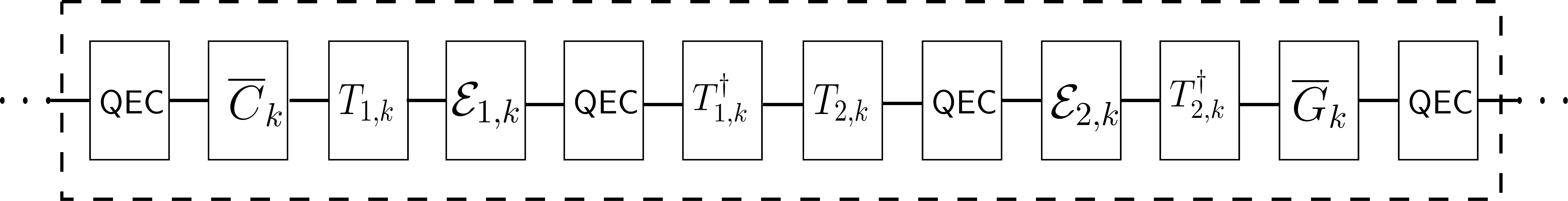}};
\node [below = 0.15cm of b] (c) {\includegraphics[scale=0.1]{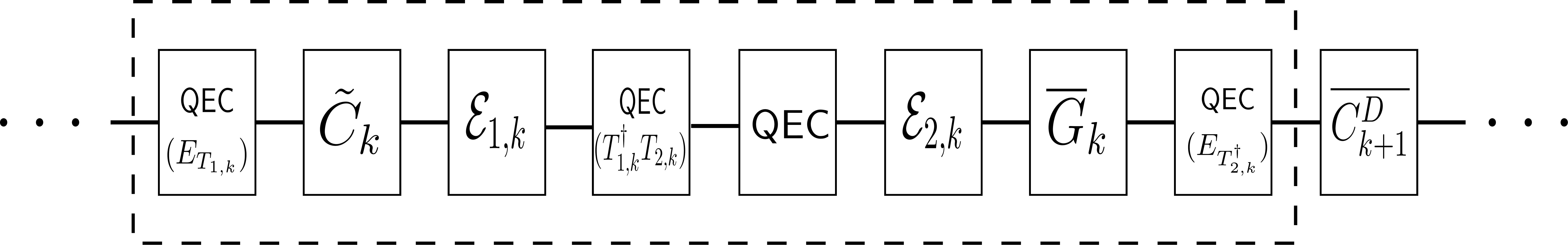}};
\node [left = 0.1cm of a] (acap) {\large (a)};
\node [left = 0.1cm of b] (bcap) {\large (b)};
\node [left = 0.1cm of c] (ccap) {\large (c)};
\end{tikzpicture}

\vspace{-0.2cm}

\caption[RC in FT algorithm]{Compiling twirling (random physical Pauli) gates into fault tolerant gadgets. Figure (a) shows the noisy gates in the $k-$th clock cycle of a fault tolerant quantum algorithm presented in the standard form prescribed in \cite{WE16}. Twirling gates are inserted in figure (b) to tailor the noise processes to Pauli errors. These gates are compiled into existing gates by replacing easy gates by their dressed versions in figure (c).}
\label{fig:rc_ftqec}
\end{figure*}
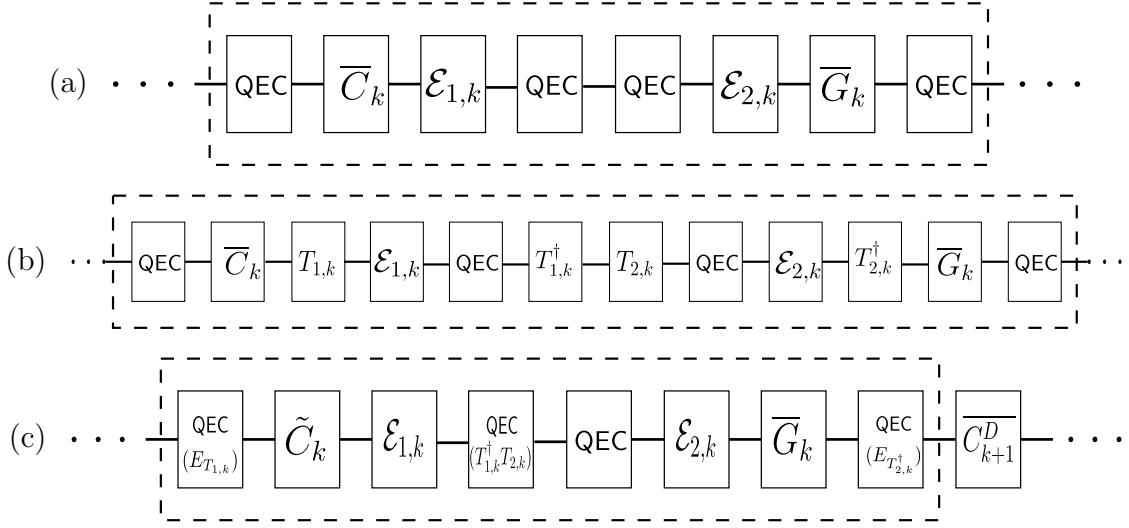

\section{Logical estimator for concatenated codes} \label{sec:uncorr}
In this section we discuss the derivation for the logical estimator $\cp$ used to predict the logical error rate under physical Pauli noise processes. A stabilizer code and a decoder pair is designed to correct a target set of errors, called \emph{correctable errors} \cite{S06,R12} $\cE_{C}$. For an $[[n,k]]$ code, $\cE_{C}$ can be partitioned into $2^{n-k}$ disjoint subsets $\cE_{C,1}, \ldots, \cE_{C,2^{n-k}}$, each of which can be identified with a unique syndrome measurement outcome. The construction of the set $\cE_{C,s}$ closely depends on the choice of a decoder. Recall that the output of a decoder on input syndrome $s$ is a Pauli recovery operator $R_{s}$, i.e., $R_{s} \in \cE_{C,s}$. A key observation to construct elements in $\cE_{C,s}$ besides $R_{s}$ is that any error of the form $R_{s} S$ where $S$ is an element of the stabilizer group is also correctable, so, $\cE_{C,s} = \{R_{s} S : S \in \cS\}$. Uncorrectable errors cause the quantum error correction scheme to fail. We adopt the notation $p_{c}$ to denote the total probability of correctable errors:
\begin{gather}
p_{c} = \sum_{E \in \cE_{C}} \chi_{E, E} \ths , \label{eq:p_c}
\end{gather}
and $p_{u}$ to denote the total probability of uncorrectable errors: $p_{u} = 1 - p_{c}$. It is easy to note that $p_{u}$ is an upper bound to the standard infidelity metric which is measured by randomized benchmarking, i.e., $r = 1 - \chi_{0,0}$:
\begin{gather}
p_{u} = r - \sum_{\substack{E\in\cE_{C} \\ E\neq \bI}} \chi_{E,E} \ths . \label{eq:uncorr_infid_nonid}
\end{gather}
In particular, for Pauli noise processes the following equations show that $p_{u}$ is exactly the average \emph{logical} infidelity $\ol{r}$.
\begin{flalign}
\ol{r} &= 1 - \sum_{\substack{E,E^{\prime} \in \cE_{C} \\ s(E) = s(E^{\prime}) \ts , \ts \ol{E} = \ol{E}^{\prime}}} \chi_{E,E^{\prime}} \label{eq:log_fidelity_chi_correctable} \\
&= r - \sum_{\substack{E, E^{\prime} \in \cE_{C} \ts , \ts E, E^{\prime} \neq \bI \\ s(E) = s(E^{\prime}) \ts, \ts \ol{E} = \ol{E}^{\prime}}} \chi_{E,E^{\prime}} \label{eq:relate_phys_infid_log_infid} \\
&= p_{u} - \sum_{\substack{E, E^{\prime} \in \cE_{C} \ts , \ts E \neq E^{\prime} \\ s(E) = s(E^{\prime}) \ts , \ts \ol{E} = \ol{E}^{\prime}}} \chi_{E,E^{\prime}} \label{eq:relate_pu_log_infid} \ths .
\end{flalign}
A detailed derivation of eq. \ref{eq:log_fidelity_chi_correctable} is presented in section \ref{sec:loginfid} of this appendix. The expressions in eqs. \ref{eq:relate_phys_infid_log_infid} and \ref{eq:relate_pu_log_infid} point out a conceptual difference between infidelity and the uncorrectable error probability. While on the one hand, $r$ accounts for the effect of only the trivial correctable error $\bI$, $p_{u}$ on the other hand captures many more degrees of freedom -- including all other correctable errors in $\cE_{C}$. Hence, we expect $r$ to be a worse predictor of the logical infidelity than $p_{u}$. 

It is generally infeasible to enumerate all the $\cO(4^{n-k})$ correctable errors for an $[[n,k]]$ stabilizer code, to compute $p_{u}$ exactly. Our logical estimator is the result of an efficient heuristic to approximate $p_{u}$, particularly for concatenated code families. Furthermore, its accuracy is provably high for uncorrelated Pauli error models.

While for concatenated codes, the number of physical qubits itself grows exponentially in the size of a code block $n$, we can exploit its encoding structure to simplify the complexity of computing $p_{u}$. However, it turns out that despite this simplification we cannot exactly compute $p_{u}$ efficiently, i.e., in time that scales polynomially in the number of physical qubits. This leads us to resort to a heuristic method for a reasonable approximation of $p_{u}$ for concatenated codes, described in the rest of this section. Here we present a method to measure and compute an approximation, denoted by $\cp(\cC^{\star}_{\ell})$, to the probability of uncorrectable errors for a concatenated code $\cC^{\star}_{\ell}$: $p_{u}(\cC^{\star}_{\ell})$. For ease of notation we also define the quantities $p_{c}(\cC^{\star}_{\ell}) = 1 - p_{u}(\cC^{\star}_{\ell})$ and $\wt{p}_{c}(\cC^{\star}_{\ell}) = 1 - \cp(\cC^{\star}_{\ell})$.

An error $E_{\ell}$ for the level $\ell$ concatenated code $\cC^{\star}_{\ell}$ can be expressed as a tensor product of Pauli errors $E_{\ell-1,i}$ for the level $\ell-1$ codes $\cC^{\star}_{\ell-1,i}$:
\begin{gather}
E_{\ell} = \bigotimes_{i=1}^{n}E_{\ell-1,i} \ths . \label{eq:L_op_L-1}
\end{gather}
Let us define $E_{\ell}$ to be a correctable pattern if the above tensor product corresponds to an encoded version of a correctable error for the code block $\cC_{\ell}$. For example, $E_{2} = \ol{X} \ot \ol{\bI}^{\ts \ot 6}$ is a correctable pattern for the $\ell=2$ concatenated Steane code since $X\ot \bI^{\ot 6}$ is a correctable error for the Steane code block.

A correctable error $E_{\ell}$ for $\cC^{\star}_{\ell}$ falls into one of the two categories: either (i) it is corrected within the lower level code-blocks $\cC^{\star}_{\ell-1, 1}, \ldots, \cC^{\star}_{\ell-1, n}$, or (ii) it has a non-trivial correction applied by the decoder of the level$-\ell$ code-block $\cC_{\ell,1}$. Let us denote the contribution to $p_{c}(\cC^{\star}_{\ell})$ from case (i) by $\Lambda$, while that from case (ii) by $\Gamma$; so that
\begin{gather}
p_{c}(\cC^{\star}_{\ell}) = \Lambda(\cC^{\star}_{\ell}) + \Gamma(\cC^{\star}_{\ell}) \ths . \label{eq:pc_contribs}
\end{gather}

Case (i) implies that each of the errors $E_{\ell-1, i}$ are correctable errors for the codes $\cC^{\star}_{\ell-1, i}$. Therefore, the total probability of correctable errors in case (i) admits a recursive definition:
\begin{gather}
\Lambda(\cC^{\star}_{\ell}) = p_{c}(\cC^{\star}_{\ell-1,1})p_{c}(\cC^{\star}_{\ell-1,2})\ldots p_{c}(\cC^{\star}_{\ell-1,n}) \ths . \label{eq:Q1}
\end{gather}

Recall that case (ii) is the total probability of non-trivial correctable patterns for $\cC^{\star}_{\ell}$, i.e.,
\begin{flalign}
\Gamma(\cC^{\star}_{\ell}) &= \sum_{E \ts\in\ts\cE_{\cC}\backslash \bI}\Pr(E_{\ell}) \ths , \label{eq:nontrivial_patterns} \\
&= \sum_{E \ts\in\ts\cE_{\cC}\backslash \bI}\Pr(E_{\ell-1,1} \otimes E_{\ell-1,2} \otimes \ldots \otimes E_{\ell-1,n}) \label{eq:patt}
\end{flalign}
where we have used the fact that each correctable error corresponds to a pattern according to eq. \ref{eq:L_op_L-1}. A logical error $E_{\ell-1,i}$ occurs on the codeblock $\cC_{\ell-1,i}$ whenever the decoder fails in correcting the physical errors in such a way that the residual effect of the physical noise process affecting the qubits of $\cC^{\star}_{\ell-1,i}$ and the recovery operation applied by the decoder results in $E_{\ell-1,i}$.
Let us denote the probability of the decoder for $\cC^{\star}_{\ell-1,i}$ to leave a residual $E_{\ell-1,i}$, conditioned on the syndrome measurements by $\Pr_{\cD}(E_{\ell-1, i}\ts | \ts s(\cC^{\star}_{\ell-1,i}))$. We can rewrite eq. \ref{eq:patt} as
\begin{flalign}
\Gamma(\cC^{\star}_{\ell}) &= \sum_{E \ts\in\ts\cE_{\cC}\backslash \bI}\ths\sum_{s(\cC^{\star}_{\ell})}\Pr(s(\cC_{\ell})s(\cC^{\star}_{\ell-1,1})\ldots s(\cC^{\star}_{\ell-1,n}))\prod_{j=1}^{n}\Pr_{\cD}(E_{\ell-1,j} | s(\cC^{\star}_{\ell-1,j})) \ths , \label{eq:Q2_residual} \\
&= \sum_{E \ts\in\ts\cE_{\cC}\backslash \bI}\ths\sum_{s(\cC^{\star}_{\ell})}\Pr(s(\cC_{\ell})|s(\cC^{\star}_{\ell-1,1})\ldots s(\cC^{\star}_{\ell-1,n}))\prod_{j=1}^{n}\Pr_{\cD}(E_{\ell-1,j} | s(\cC^{\star}_{\ell-1,j})) \Pr(s(\cC^{\star}_{\ell-1,j})) \ths , \label{eq:Q2_residual_conditional}
\end{flalign}
where $\Pr(s(\cC_{\ell}) | s(\cC^{\star}_{\ell-1,1})\ldots s(\cC^{\star}_{\ell-1,n}))$, is the conditional probability of measuring the syndrome outcomes $s(\cC_{\ell})$ on the codeblock $\cC_{\ell}$ when the outcomes on the lower level code blocks $\cC^{\star}_{\ell-1,1}, \ldots, \cC^{\star}_{\ell-1,n}$ are $s(\cC^{\star}_{\ell-1,1}),\ldots, s(\cC^{\star}_{\ell-1,n})$, respectively. Equivalently,
\begin{gather}
\Pr(s(\cC_{\ell}) | s(\cC^{\star}_{\ell-1,1})\ldots s(\cC^{\star}_{\ell-1,n})) = \Pr(s(\cC_{\ell})|\cE^{s(\cC^{\star}_{\ell-1,1})}_{\ell-1,1}\ldots \cE^{s(\cC^{\star}_{\ell-1,n})}_{\ell-1,n})) \ths \label{eq:exact_conditional} .
\end{gather}
A major hurdle in computing $\Gamma$ using eq. \ref{eq:Q2_residual_conditional} is the sum over an exponentially large set of syndrome outcomes for the concatenated code. To circumvent this difficultly, we will apply an efficient heuristic to approximate the probability in eq. \ref{eq:exact_conditional}. In essence, we will replace the conditional channel $\cE^{s(\cC^{\star}_{\ell-1,i})}_{\ell-1,i}$ by the average logical channel $\hat{\cE}_{\ell-1,i}$, which is defined as
\begin{gather}
\hat{\cE}_{\ell-1,i} = \sum_{s(\cC_{\ell-1,i})}\Pr(s(\cC_{\ell-1,i}))\cE^{s(\cC_{\ell-1,i})}_{\ell-1,i}\left[\hat{\cE}_{\ell-2,1} \otimes \ldots \otimes \hat{\cE}_{\ell-2,n}\right] \label{eq:coarse_grained} \ths .
\end{gather}
Note that $\hat{\cE}_{0,j}$ is the physical noise model while $\hat{\cE}_{1,j}$ is the exact average logical channel $\ol{\cE}_{1,j}$. However, in general for $\ell \geq 2$, $\hat{\cE}_{\ell}$ is a coarse-grained approximation for the exact average logical channel $\ol{\cE}_{\ell}$. In other words, $\hat{\cE}_{\ell-1,i}$ is computed using the knowledge of the syndrome bits measured only at level $\ell-1$, while assuming the noise model: $\hat{\cE}_{\ell-2,1} \otimes \ldots \otimes \hat{\cE}_{\ell-2,n}$, that accounts for the average effect of all syndrome measurements at lower levels.

Replacing the conditional channel $\cE^{s(\cC^{\star}_{\ell-1,i})}_{\ell-1,i}$ in eq. \ref{eq:Q2_residual_conditional} by the average channel $\hat{\cE}_{\ell-1,i}$ defined in eq. \ref{eq:coarse_grained} allows us to approximate $\Gamma$ by $\wt{\Gamma}$ defined as follows:
\begin{flalign}
\wt{\Gamma}(\cC^{\star}_{\ell}) &= \sum_{E \ts\in\ts\cE_{\cC}\backslash \bI}\ths\sum_{s(\cC_{\ell})}\sum_{s(\cC^{\star}_{\ell-1,1})}\ldots\sum_{s(\cC^{\star}_{\ell-1,n})}\Pr(s(\cC_{\ell})|\hat{\cE}_{\ell-1,1}\ldots \hat{\cE}_{\ell-1,n}))\prod_{j=1}^{n}\Pr_{\cD}(E_{\ell-1, j}\ts | \ts \hat{\cE}_{\ell-1,j})\Pr(s(\cC^{\star}_{\ell-1,j})) \label{eq:crucial_approx} \ths , \\
&= \sum_{E \ts\in\ts\cE_{\cC}\backslash \bI}\ths\prod_{j=1}^{n}\Pr_{\cD}(E_{\ell-1, j}\ts | \ts \hat{\cE}_{\ell-1,j})) \ths . \label{eq:Q2_approx}
\end{flalign}
Denote $\cR(s(\cC_{\ell-1,i}), P)$ to be the set of $n-$qubit errors on which a lookup table decoder for the code block $\cC_{\ell-1,i}$ leaves a residual logical error $P$ when the error syndrome $s(\cC_{\ell-1,i})$ is encountered. Now $\Pr_{\cD}(E_{\ell-1, i}\ts | \ts \hat{\cE}_{\ell-1,j}))$ can be computed recursively:
\begin{gather}
\Pr_{\cD}(E_{\ell-1, i}\ts | \ts \hat{\cE}_{\ell-1,i})) = \sum_{Q \in \cR(s(\cC_{\ell-1,i}), E_{\ell-1,i})} \prod_{j=1}^{n}\Pr_{\cD}(Q_{\ell-2, j}\ts | \ts \hat{\cE}_{\ell-2,j})) \ths . \label{eq:residual_recursion}
\end{gather}
Note that the probability of leaving a residual error at level 0 is simply specified by the physical noise model, i.e., $\Pr_{\cD}(P | \hat{\cE}_{0,j})$ is the probability of the Pauli error $P$ on the physical qubit $j$. This concludes the method to efficiently compute $\wt{\Gamma}$, an approximation to $\Gamma$.

Recall that the total probability of correctable errors is given by eq. \ref{eq:pc_contribs}. An approximation to $p_{c}(\cC^{\star}_{\ell})$, is given by
\begin{gather}
\wt{p}_{c}(\cC^{\star}_{\ell}) = \wt{\Lambda}(\cC^{\star}_{\ell}) + \wt{\Gamma}(\cC^{\star}_{\ell}) \ths , \label{eq:pc_approx}
\end{gather}
where $\wt{\Gamma}$ defined in eq. \ref{eq:Q2_approx} while $\wt{\Lambda}$ is defined in a similar fashion to eq. \ref{eq:Q1}:
\begin{gather}
\wt{\Lambda}(\cC^{\star}_{\ell}) = \wt{p}_{c}(\cC^{\star}_{\ell-1,1})\wt{p}_{c}(\cC^{\star}_{\ell-1,2})\ldots \wt{p}_{c}(\cC^{\star}_{\ell-1,n}) \ths . \label{eq:Q1_approx}
\end{gather}
Using the approximation in eq. \ref{eq:pc_approx}, we can efficiently estimate the logical estimator $\cp$ for concatenated codes.

\section{Approximation quality for the uncorrectable error probability} \label{sec:pu_quality}
In this section, we will quantity the accuracy of the approximating the uncorrectable error probability using $\cp$ for concatenated codes. For simplicity, we will assume that the code-blocks in the concatenated code are all identical, and equal to a $[[n,1,d]]$ quantum error correcting code, with $d\geq 3$. Recall that the distance of a level $\ell$ concatenated code scales as $d^{\ell}$. We will use $t_{\ell} = \lfloor(d^{\ell}+1)/2\rfloor$ to denote the Hamming weight of the smallest uncorrectable error. Recall that $\cp$ is defined recursively as the sum of two quantities: $\wt{Q}_{1}$ and $\wt{Q}_{2}$. We will use $\delta_{\ell}$ to denote the inaccuracy in computing $p_{u}$ for a level $\ell$ concatenated code:
\begin{gather}
\delta_{\ell} = |p_{u}(\cC^{\star}_{\ell,1}) - \cp(\cC^{\star}_{\ell,1})| \ths , \label{eq:delta_pu}
\end{gather}
and $\gamma_{\ell}$ to denote the inaccuracy in computing $\Gamma$:
\begin{gather}
\gamma_{\ell} = |\wt{\Gamma}(\cC^{\star}_{\ell}) - \Gamma(\cC^{\star}_{\ell})| \ths . \label{eq:delta_Q2}
\end{gather}
Then it follows that
\begin{gather}
\delta_{\ell} \leq n \delta_{\ell-1} + \gamma_{\ell} \ths . \label{eq:error_scaling}
\end{gather}
The most important ingredient in computing $\delta_{\ell}$ is $\gamma_{\ell}$, defined in eq. \ref{eq:delta_Q2}. For simplicity we will compute $\gamma_{\ell}$ for the i.i.d depolarizing error model. However, for generic i.i.d Pauli error models, we can replace the depolarizing rate $p$ in our analysis by the physical infidelity of the single qubit error model, $r_{0}$. The extension to correlated Pauli error models remains unclear.

An i.i.d application of the depolarizing channel on $n-$qubits can be described by
\begin{gather}
\cE(\rho) = \sum_{P \ts \in \ts \cP_{n}}\chi_{P,P} P\ts \rho \ts P \ths , \nonumber \\
\text{such that } \chi_{P,P} = (1-p)^{n - |P|}\left(\dfrac{p}{3}\right)^{|P|} \ths , \label{eq:chi_dp}
\end{gather}
where $\cP_{n}$ is the $n-$qubit Pauli group, $0\leq p\leq 1$ is the depolarizing rate and $|P|$ is the Hamming weight of the Pauli error $P$. In this case, we will show that
\begin{gather}
\gamma_{\ell} = \cO(n^{\ell-1} p^{t_{\ell-1}+2}) \ths , \label{eq:Q2_approx_dp}
\end{gather}
for a level $\ell$ concatenated code.

Combining eq. \ref{eq:Q2_approx_dp} with eq. \ref{eq:error_scaling}, we arrive at an expression for $\delta_{\ell}$:
\begin{gather}
\delta_{\ell} = \cO(n^{\ell-1} p^{2+\left\lfloor (d+1)/2\right\rfloor}) \ths , \label{eq:delta_pu_scaling}
\end{gather}
where $d$ is the distance of a code block.

In the rest of this section, we will derive eq. \ref{eq:Q2_approx_dp}. Recall that eq. \ref{eq:Q2_approx} outlines the approximation made by the heuristic to compute $\Gamma(\cC^{\star}_{\ell})$. It involves replacing the knowledge of conditional channels $\cE^{s}_{\ell-1,j}$ by the average channel, $\hat{\cE}_{\ell-1,j}$. We will prove the scaling in eq. \ref{eq:Q2_approx_dp} two steps. First, is an observation that
\begin{gather}
\prod_{j=1}^{n}\Pr_{\cD}(E_{\ell-1,j} | \hat{\cE}_{\ell-1,j}) = \cO(p^{t_{\ell-1}}) \ths . \label{eq:residual_order}
\end{gather}
This follows from the fact that at least one of the errors $E_{\ell-1,j}$ in the error pattern $E_{\ell-1,1}\otimes \ldots \otimes E_{\ell-1,n}$ must be non-identity. Note that a non-identity logical error is left as a residual when the decoder for the subsequent lower level fails. Such an event will not occur for errors whose weight is below $t_{\ell-1}$.

Second, by showing that
\begin{flalign}
\Pr(s(\cC_{\ell,i}) | \cE^{s(\cC^{\star}_{\ell-1,1})}_{\ell-1,1}\ldots \cE^{s(\cC^{\star}_{\ell-1,n})}_{\ell-1,n}) &\prod_{i=1}^{n}\Pr(s(\cC^{\star}_{\ell-1,j})) = \nonumber \\
&\Pr(s(\cC_{\ell,i}) | \hat{\cE}_{\ell-1,1}\ldots \hat{\cE}_{\ell-1,n}) \prod_{i=1}^{n}\Pr(s(\cC^{\star}_{\ell-1,j})) + \cO(n^{\ell-1} p^{2}) \ths . \label{eq:approx_quality}
\end{flalign}
Recall from eq. \ref{eq:coarse_grained} that the average channel $\hat{\cE}_{\ell,i}$ is defined recursively in terms of $\hat{\cE}_{\ell-1,j}$. While the term corresponding to $s(\cC_{\ell,i}) = 0$ describes the effect of stabilizers on the input state, the other terms include the effect of non-trivial errors. Note that the a non-trivial error $E_{\ell}$ has weight at least $t_{\ell-1}$, equal to the weight of the smallest uncorrectable error of the concatenated code $\cC^{\star}_{\ell-1,j}$. Carrying this idea from level $\ell-1$ to level $1$, we find:
\begin{flalign}
\hat{\cE}_{\ell,i} &= \cE^{s(\cC_{\ell,i}) = 0}_{\ell,i}\left[\hat{\cE}_{\ell-1,1} \otimes \ldots \otimes \hat{\cE}_{\ell-1,n}\right] + \cO(p^{t_{\ell-1}}) \ths , \label{eq:avgL_order} \\
&= \left(\hat{\cE}_{\ell-1,1} \otimes \ldots \otimes \hat{\cE}_{\ell-1,n}\right) + \cO(p^{t_{\ell-2}}) \ths , \label{eq:avgL_order_proj} \\
&= \left(\hat{\cE}_{1,1} \otimes \ldots \otimes \hat{\cE}_{1,n^{\ell-1}}\right) + \cO(p^{t_{1}}) \ths , \label{eq:avgL1_order_proj}
\end{flalign}
where in eq. \ref{eq:avgL_order_proj} we have used the fact that the leading contribution to the conditional channel for the trivial syndrome, is the physical channel itself. Equation \ref{eq:avgL1_order_proj} describes the recursion until level $\ell=1$ where $\hat{\cE}_{1,j} = \ol{\cE}_{1,j}$.

Recall that the conditional channel for an error-syndrome $s(\cC^{\star}_{\ell-1,i})$,
\begin{gather}
\cE^{s(\cC^{\star}_{\ell-1,i})}_{\ell-1,i} = \cE^{s(\cC_{\ell-1,i})s(\cC_{\ell-2,1})\ldots s(\cC_{\ell-2,n})\ldots s(\cC_{1,1})\ldots s(\cC_{1,n^{\ell-1}})}_{\ell-1,i} \label{eq:conditional_synd_expanded} \ths ,
\end{gather}
is defined by applying quantum error correction routines corresponding to the syndrome outcomes in the respective code-blocks of $\cC^{\star}_{\ell-1,i}$. Note that an error is detected (by means of a non-trivial syndrome outcome) in a code block at level $\ell$ when the decoder operating on the code block at level $\ell-1$ leaves a non-trivial residue. Hence, for a leading order analysis, we will consider conditional channels that correspond to trivial syndromes in all the code-blocks except for those at level one, i.e., $s(\cC_{\ell,i}) = 0$ for all $\ell > 1$ in eq. \ref{eq:conditional_synd_expanded}. In other words, we will consider errors that are corrected within the code blocks in level one:
\begin{gather}
\cE^{s(\cC_{\ell-1,i})= 0,\ts s(\cC_{\ell-2,1}) = 0, \ts \ldots , \ts s(\cC_{\ell-2,n}) = 0, \ts \ldots, \ts s(\cC_{1,1})\ldots s(\cC_{1,n^{\ell-1}})}_{\ell-1,i} = \cE^{s(\cC_{1,1})}_{1,1} \otimes \ldots \otimes \cE^{s(\cC_{1,n^{\ell-1}})}_{1,n^{\ell-1}} + \cO(p^{t_{1}}) \ths . \label{eq:conditional_channel_trivial}
\end{gather}

Using eqs. \ref{eq:avgL1_order_proj} and \ref{eq:conditional_channel_trivial}, we note that the quality of the approximation in eq. \ref{eq:approx_quality} can be bounded as follows:
\begin{flalign}
&\left(\Pr(s(\cC_{\ell}) | \cE^{s(\cC_{1,1})}_{1} \ldots \cE^{s(\cC_{1,n^{\ell-1}})}_{1}) - \Pr(s(\cC_{\ell}) | \hat{\cE}_{1,1}\ldots \hat{\cE}_{1,n^{\ell-1}})\right) \prod_{j=1}^{n^{\ell-1}}\Pr(s(\cC_{1,j})) \nonumber \\
&= \Tr\left[\Pi_{s(\cC_{\ell})} \cdot \left((\cE^{s(\cC_{1,1})}_{1}\otimes\ldots \otimes\cE^{s(\cC_{1,n^{\ell-1}})}_{1})(\rho) - (\hat{\cE}_{1,1}\otimes\ldots \otimes\hat{\cE}_{1,n^{\ell-1}})(\rho)\right)\right] \prod_{j=1}^{n^{\ell-1}}\Pr(s(\cC_{1,j})) \ths , \label{eq:diff_trace} \\
&= \sum_{i}\left[\left(\chi^{s(\cC_{1,1})}_{1,1} \otimes \ldots \otimes \chi^{s(\cC_{1,n^{\ell-1}})}_{1,n^{\ell-1}}\right)_{i,i} - \left(\hat{\chi}_{1,1} \otimes \ldots \otimes \hat{\chi}_{1,n^{\ell-1}}\right)_{i,i}\right] \ths \Tr\left[\Pi_{s(\cC_{\ell})} \cdot P_{i}\rho P_{i}\right] \prod_{j=1}^{n^{\ell-1}}\Pr(s(\cC_{1,j})) \ths \label{eq:diff_chi} , \\
&\leq n^{\ell-1}\max_{s \ts \in \ts \bZ^{n-k}_{2}}||\chi^{s}_{1} - \hat{\chi}_{1}||_{\infty}\Pr(s) \ths , \label{eq:diff_chi_max}
\end{flalign}
where $\chi^{s}_{1}$ refers to the chi matrix of the conditional channel $\cE^{s}_{1}$ while $\hat{\chi}_{1}$ refers to the chi matrix of the average channel $\hat{\cE}_{1}$. In eq. \ref{eq:diff_chi_max}, we have used the matrix norm $||A||_{\infty}$ to refer to the maximum absolute value in the matrix.

To establish the scaling in eq. \ref{eq:approx_quality} it remains to show that
\begin{gather}
\max_{s\ts\in\ts\bZ_{2}^{n-k}}||\chi^{s}_{1} - \hat{\chi}_{1}||_{\infty}\Pr(s) = \cO(p^{2}) \ths . \label{eq:probsynd_order}
\end{gather}
Recall that the effective channel for a given syndrome $s$: $\cE^{s}_{1}$, describes the composite effect of the physical noise process and quantum error correction conditioned on the measurement outcome $s$. Comparing eq. \ref{eq:average_eff_state_chi_no_projs} to the general form in eq. \ref{eq:logical_chi_channel}, we find an expression similar to eq. \ref{eq:chi_lm_log_phys}:
\begin{gather}
\left[\chi^{s}_{1}\right]_{i,i} = \dfrac{1}{\Pr(s)}\sum_{\substack{E \in \cE_{\cC} \\ s(E) = s}} \chi_{\ol{P}_{i} E, \ol{P}_{i}E} \ths . \label{eq:log_chi_expr}
\end{gather}

For the specific case of the depolarizing channel in \ref{eq:chi_dp} we can express $\left[\chi^{s}_{1}\right]_{i,i}$, $\Pr(s)$ and $\hat{\chi}_{i,i}$ as polynomials in the depolarizing rate $p$:
\begin{flalign}
\left[\chi^{s}_{1}\right]_{i,i} &= \dfrac{1}{\Pr(s)}\sum_{w=1}^{n}A^{s}_{i,w}(1-p)^{n-w}\left(\dfrac{p}{3}\right)^{w} \ths , \label{eq:log_chi_A} \\
\Pr(s) &= \sum_{i}\sum_{w=1}^{n}A^{s}_{i,w}(1-p)^{n-w}\left(\dfrac{p}{3}\right)^{w} \ths , \\
\hat{\chi}_{i,i} &= \sum_{s}\sum_{w=0}^{n}A^{s}_{i,w}(1-p)^{n-w}\left(\dfrac{p}{3}\right)^{w} \ths , 
\end{flalign}
where $A^{s}_{i,w}$ is the number of Pauli errors $Q$ of Hamming weight $w$ on which the action of the decoder leaves a residual logical error $\ol{P}_{i}$. In other words, $Q = \ol{P}_{i} R_{s} S$ where $R_{s}$ is the recovery operation prescribed by the decoder for the error-syndrome $s$ and $S$ is any stabilizer. We can use two simple facts about errors to simplify the coefficients $A^{s}_{i,w}$. First, since the only error of Hamming weight zero is the identity which has $s=0$, we find $A^{s}_{i,0} = \delta_{s,0}\delta_{i,0}$. Second, since all errors of Hamming weight up to $\lfloor(d-1)/2\rfloor$ are correctable, we find $A^{s}_{i,w} = \delta_{i,0}A^{s}_{0,w}$ for all $w \leq \lfloor(d-1)/2\rfloor$. Using these simplifications,
\begin{flalign}
\left[\chi^{s}_{1}\right]_{i,i} &= \dfrac{1}{\Pr(s)}\left[(1-p)^{n-1}\left(\dfrac{p}{3}\right)A^{s}_{0,1} + \cO(n^{2}p^{2})\right] \ths , \\
\Pr(s) &= A^{s}_{0,1}(1-p)^{n-w}\left(\dfrac{p}{3}\right) + \cO(n^{2}p^{2}) \ths , \\
\hat{\chi}_{i,i} &= \delta_{i,0}(1-p)^n + 3 n (1-p)^{n-1}\left(\dfrac{p}{3}\right)\delta_{i,0} + \cO(n^{2}p^{2})  \ths .
\end{flalign}
It is now straightforward to see that eq. \ref{eq:probsynd_order} follows from the above set of equations.

In summary, this section establishes that the approximation used by the heuristic to compute $\cp(\cC^{\star}_{\ell,1})$, is accurate to $\cO(n^{\ell+1}p^{2+\lfloor(d+1)/2\rfloor})$ for the i.i.d depolarizing physical error model with error rate $p$. To get a sense for this approximation quality, we can plug in relevant numbers for an i.i.d Pauli error model and level-$2$ concatenated Steane code: $p = 10^{-3}, n = 7, \ell = 2, d = 3$. Numerical simulations of quantum error correction yield an estimate of the logical infidelity given by $4.2 \times 10^{-9}$. The analytical bound suggests that the logical estimator derived from the our heuristic method agrees with the logical infidelity up to $\cO(10^{-11})$. However, the scaling suggests that the heuristic may not be not accurate for large codes in the high noise regime. Nonetheless we have strong numerical evidence to support that the logical estimator predicts the functional form of logical infidelity.

\section{Time complexity of computing $\cp$ for concatenated codes} \label{sec:pu_time_app}
Recall that $\cp = 1 - \wt{p}_{c}(\cC^{\star}_{\ell})$, where $\wt{p}_{c}(\cC^{\star}_{\ell})$ is an approximation to the total probability of correctable errors. We will analyze the time complexity of the technique described in section \ref{sec:uncorr} to compute $\wt{p}_{c}(\cC^{\star}_{\ell})$ here.

Note that $\wt{p}_{c}(\cC^{\star}_{\ell}) = \wt{\Lambda} + \wt{\Lambda}$ where both $\wt{\Lambda}$ and $\wt{\Gamma}$ are defined recursively. So, if computing $\wt{p}_{c}(\cC^{\star}_{\ell})$ takes time $\tau_{\ell}$ and computing $\wt{\Gamma}$ takes time $\kappa_{\ell}$, we have
\begin{gather}
\tau_{\ell} = n \ts \tau_{\ell-1} + \kappa_{\ell} \ths . \label{eq:pc_time_recurrence}
\end{gather}
The recurrence relation in eq. \ref{eq:Q2_approx} for computing $\wt{\Gamma}(\cC^{\star}_{\ell})$ implies
\begin{flalign}
\kappa_{\ell} &= 4n \ths \kappa_{\ell-1} + \cO(4^n) \ths , \label{eq:recurrence_Q} \\
&= \cO(4^{n+\ell}\ths n^{\ell}) \ths . \label{eq:solution_Q}
\end{flalign}
Using the above solution in eq. \ref{eq:pc_time_recurrence}, we find that
\begin{gather}
\tau_{\ell} = \cO(4^{n+\ell} \ths n^{\ell}) \ths . \label{eq:eq:solution_T}
\end{gather}

\section{Predictability results for coherent errors}
\label{sec:results_coherent}
Numerical results presented in section \ref{sec:numerics} highlight the predictive power of the tools developed in this work with respect to the standard error-metrics, under random CPTP maps. Although CPTP maps encompass a wide range of physical noise processes, our method of generating random CPTP maps does not draw attention to an important class of noise processes -- coherent errors -- a special case of CPTP maps under which the evolution of a qubit is described by a unitary matrix. They occur due to imperfect control quantum devices and calibration errors \cite{MB14,MAB20}. Various methods such as dynamical decoupling \cite{YWL10,PGL13}, designing pulses using optimal control theory \cite{KRKS05} and machine learning approaches \cite{NBS19} are used to mitigate these errors. However, each of these methods have their shortcomings and unitary errors continue to form a major part of the total error budget \cite{GD17,HDF18,BEK18}. The methods presented in this paper will be particularly advantageous in these cases.

In this section we highlight the predictive power of our tool, over standard error metrics, under different coherent noise processes. We choose a simple class of coherent errors modeled by an unknown unitary $U_{i}$ on each physical qubit $i$, of the form $U = e^{-i \frac{\pi}{2}\delta \hat{n}\cdot \vec{\sigma}}\ths$, where $\delta$ is the angle of rotation about an axis $\hat{n}$ on the Bloch sphere. With a slight loss of generality, we will consider $n-$qubit unitary errors of the form $\otimes_{i=1}^{n} U_i$. We control the noise strength by rotation angles $\delta_i$ drawn from a normal distribution of mean and variance equal to $\mu_\delta$ where $10^{-3}\leq \mu_\delta\leq 10^{-1}$.

Figure \ref{fig:pred_coherent} shows that logical error rates vary over several orders of magnitudes across coherent errors with noise strength as measured by standard error-metrics such as infidelity and the diamond distance. In contrast, our tools provide an accurate prediction using the logical estimator developed in section \ref{sec:uncorr}. Moreover, we observe a drastic gain in in predictability using our tools for this case of unitary errors, when compared to CPTP maps in figure \ref{fig:pred} of the main text.
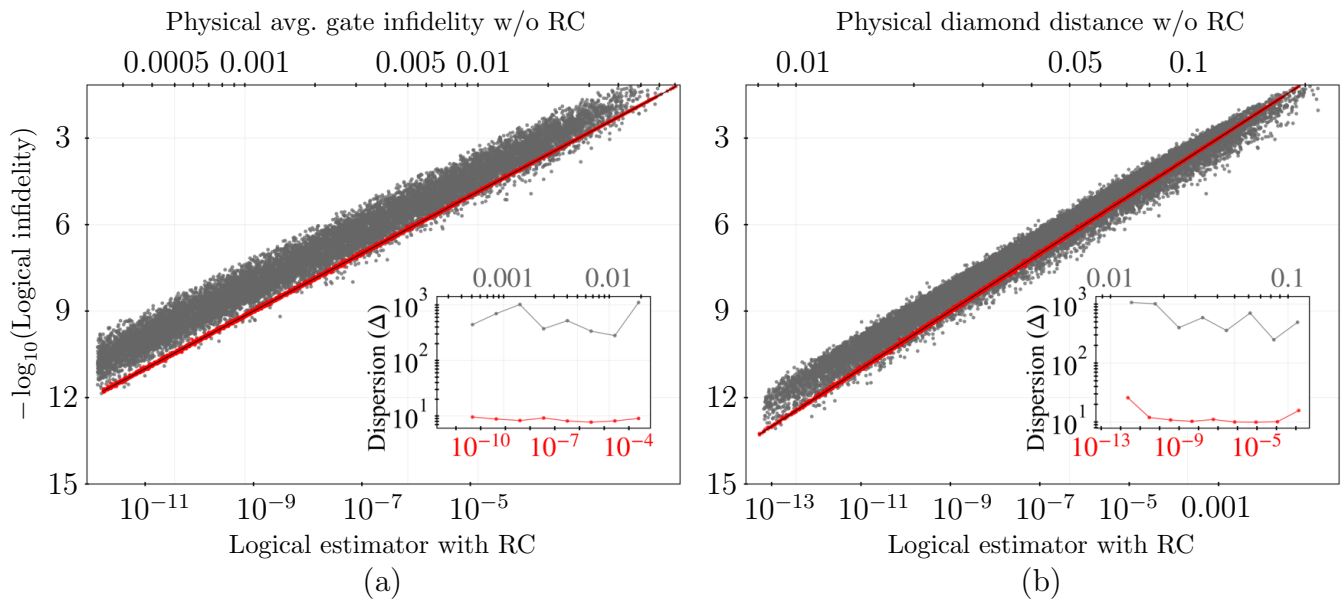
\begin{figure*}
\begin{tikzpicture}
\tikzstyle{every node}=[font=\large]
\node (a) at (-6,0) {\includegraphics[scale=0.6]{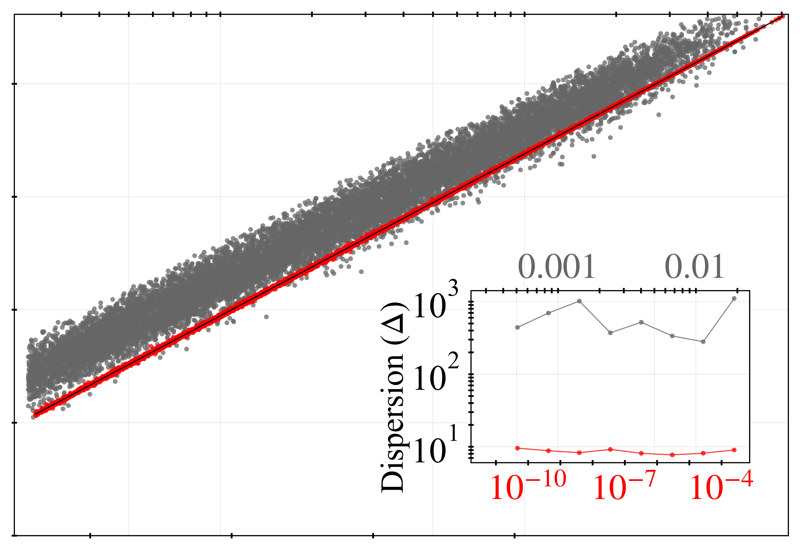}};
\node [above left = -1.2cm and -0.15cm of a] (ayt-3) {$3$};
\node [below = 0.65cm of ayt-3] (ayt-6) {$6$};
\node [below = 0.65cm of ayt-6] (ayt-9) {$9$};
\node [below = 0.65cm of ayt-9] (ayt-12) {$12$};
\node [below = 0.65cm of ayt-12] (ayt-15) {$15$};
\node [above left = -0.2cm and -2cm of a] (ait-5_4) {$0.0005$};
\node [right = -0.1cm of ait-5_4] (ait-3) {$0.001$};
\node [right = 0.9cm of ait-3] (ait-5_3) {$0.005$};
\node [right = -0.1cm of ait-5_3] (ait-2) {$0.01$};
\node [below left = -0.23cm and -1.75cm of a] (aut-11) {$10^{-11}$};
\node [right = 0.37cm of aut-11] (aut-9) {$10^{-9}$};
\node [right = 0.37cm of aut-9] (aut-7) {$10^{-7}$};
\node [right = 0.37cm of aut-7] (aut-5) {$10^{-5}$};
\node [above left = 0.27cm and -7cm of a] (atopx) {\normalsize Physical avg.~gate infidelity w/o RC};
\node [below = 0.3cm of a] (abottomx) {\normalsize Logical estimator with RC};
\node [rotate=90, above left = -0.7cm and 0.3cm of a] (aylabel) {\normalsize $-\log_{10}(\text{Logical infidelity})$};
\node [below = -0.1cm of abottomx] (acap) {(a)};
%
\node [right = 0.4cm of a] (b) {\includegraphics[scale=0.6]{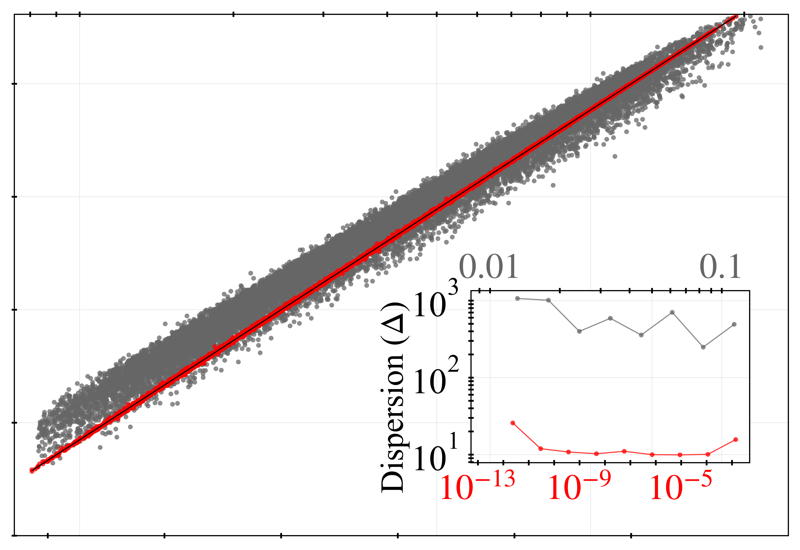}};
%
\node [above left = -1.2cm and -0.15cm of b] (byt-3) {$3$};
\node [below = 0.65cm of byt-3] (byt-6) {$6$};
\node [below = 0.65cm of byt-6] (byt-9) {$9$};
\node [below = 0.65cm of byt-9] (byt-12) {$12$};
\node [below = 0.65cm of byt-12] (byt-15) {$15$};
\node [above left = -0.2cm and -1.5cm of b] (ait-2) {$0.01$};
\node [right = 2.65cm of ait-2] (ait-5_2) {$0.05$};
\node [right = 0.65cm of ait-5_2] (ait-1) {$0.1$};
\node [below left = -0.23cm and -1.3cm of b] (but-13) {$10^{-13}$};
\node [right = 0.06cm of but-13] (but-11) {$10^{-11}$};
\node [right = 0.06cm of but-11] (but-9) {$10^{-9}$};
\node [right = 0.06cm of but-9] (but-7) {$10^{-7}$};
\node [right = 0.06cm of but-7] (but-5) {$10^{-5}$};
\node [right = 0.06cm of but-5] (but-3) {$0.001$};
\node [above left = 0.27cm and -7cm of b] (btopx) {\normalsize Physical diamond distance w/o RC};
\node [below = 0.3cm of b] (bbottomx) {\normalsize Logical estimator with RC};
\node [below = -0.1cm of bbottomx] (bcap) {(b)};
\end{tikzpicture}

\vspace{-0.2cm}

\caption[predictability]{Predicting the performance of level-2 concatenated Steane code under unitary errors. Figures (a) and (b) compare the predictive powers of our tool (red) with the standard error metrics: infidelity and the diamond distance, respectively, under an ensemble of 16000 random unitary channels. These are similar to figures 1(a) and 1(b) in the main text. The dispersion in the scatter corresponding to a metric ($\Delta$ in the insets) is indicative of its predictive power. The gains in predictability offered by our tool is drastic for the above case of unitary errors when compared to CPTP maps.}
\label{fig:pred_coherent}
\end{figure*}

\section{Importance sampling} \label{sec:importance}
A straightforward technique to estimate the logical error rate involves sampling syndrome outcomes according to the syndrome probability distribution for a quantum error correcting code and a physical noise process pair. However, there are serious drawbacks to this sampling method, due to the presence of rare syndromes -- whose probability is typically less than the inverse number of syndrome samples. A detailed account of this can be found in \cite{IP17} and in section 3.3 of \cite{I18}. We briefly review the technique here for completeness.

\begin{figure}[t]
\begin{center}
\begin{subfigure}{0.48\textwidth}
\centering
\includegraphics[scale=0.11]{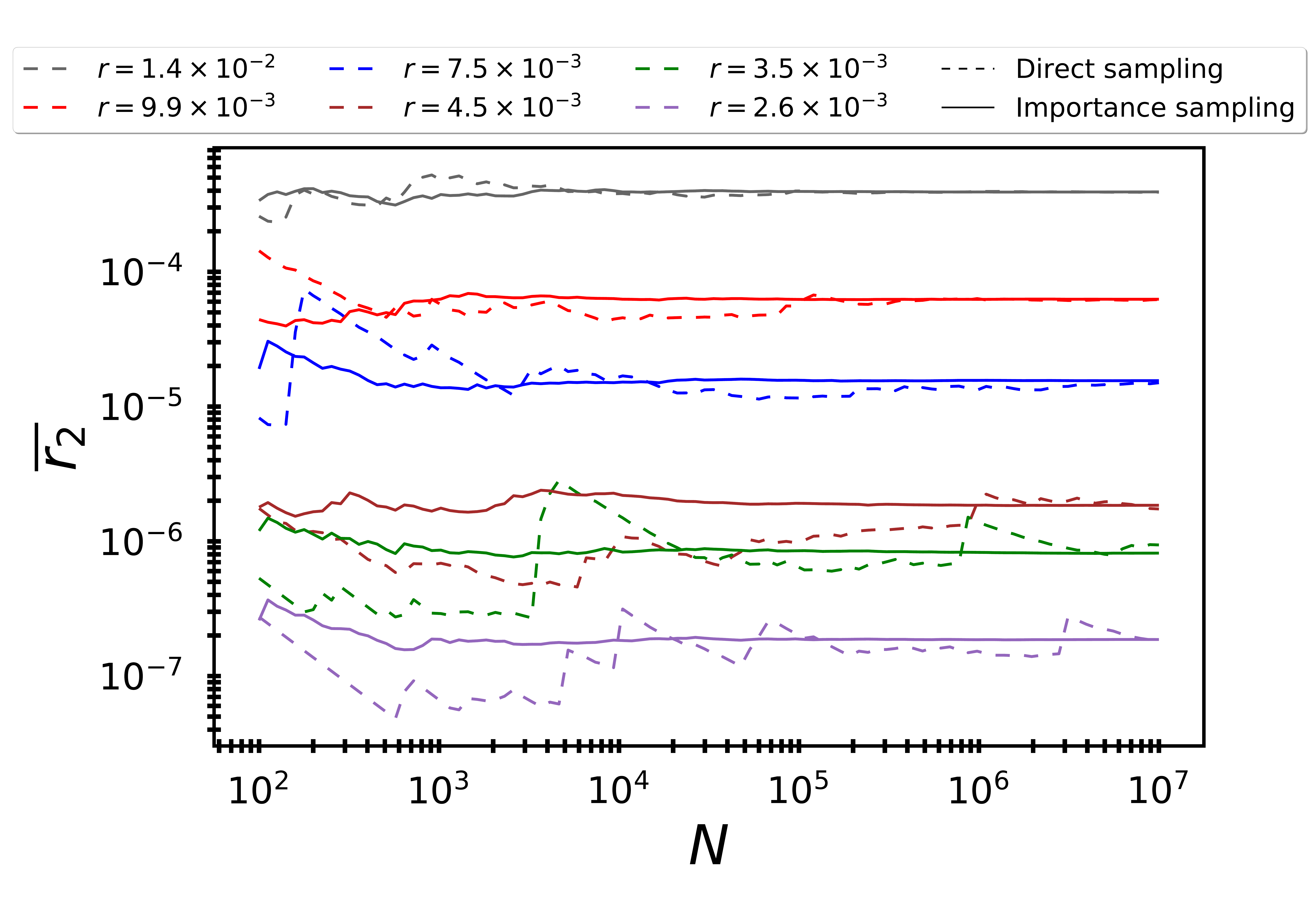}

\hspace{1.1cm} (a)
\label{fig:mc_cptp}
\end{subfigure}
\hspace{0.3cm}
\begin{subfigure}{0.48\textwidth}
\centering
\includegraphics[scale=0.11]{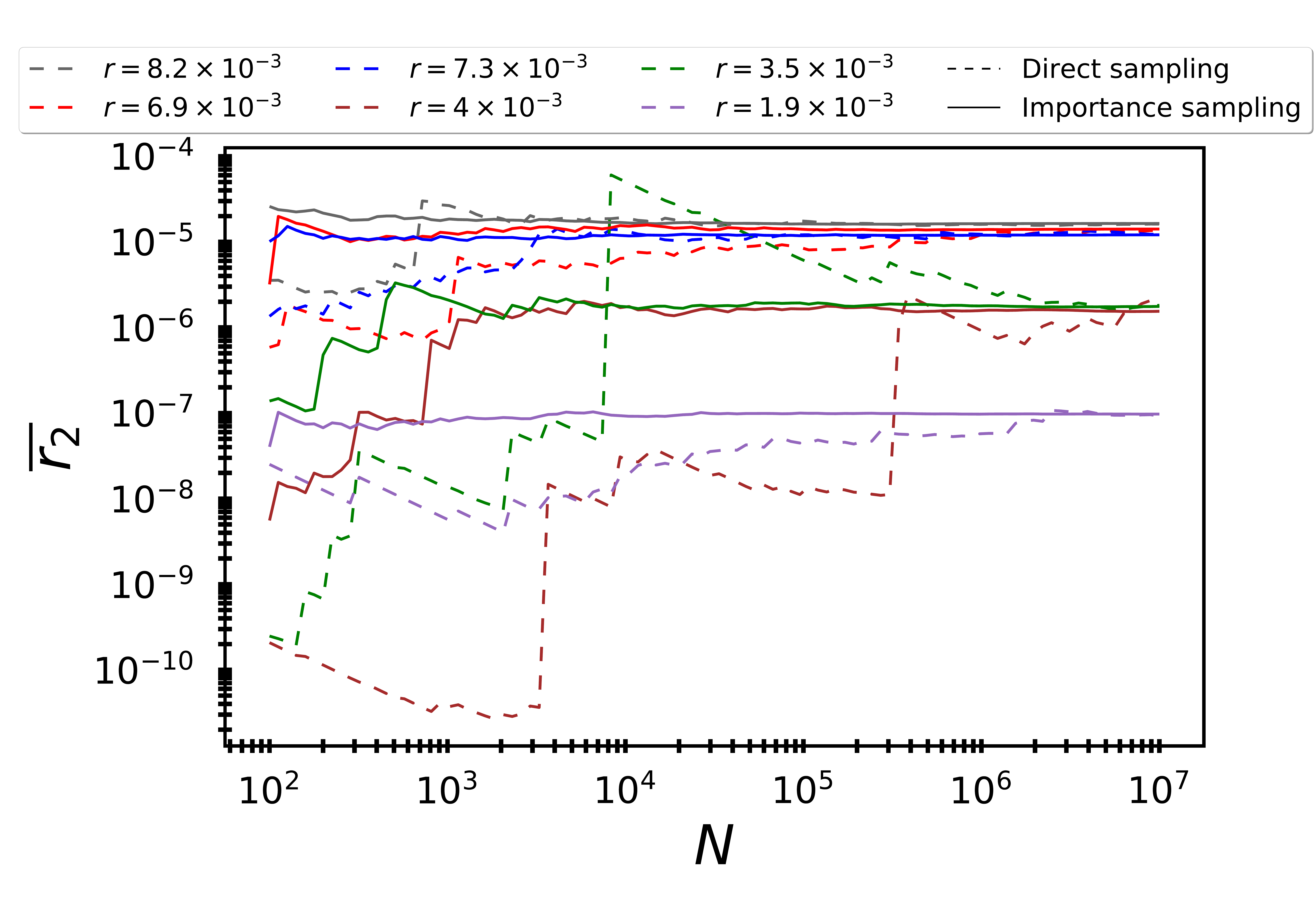}

\hspace{1.3cm} (b)
\label{fig:mc_ia}
\end{subfigure}
\caption{The above figures highlight the rapid convergence rate of the importance sampler as compared to the direct sampler, under CPTP noise processes in Fig.~6(a) and coherent errors in Fig.~6(b). Each trend line in the figures is associated to a physical noise rate. While different colors are used to identify different physical error rates, the solid and dashed lines are used to distinguish between the sampling techniques. Note that while the direct sampler takes a large number to syndrome samples to provide a reliable estimate of $r(\ol{\cE}_{\ell})$, the importance sampler achieves this task with far lesser syndrome samples. The speedup offered by importance sampling is quite drastic. The case for $r = 4\times 10^{-3}$ in Fig.~6(b) is a good example. The direct sampler shows signs of convergence around $10^{7}$ syndrome samples, whereas the importance sampler converges with just $10^{4}$ samples. Notice however that with only $10^{4}$ samples, the direct sampler underestimates $r(\ol{\cE}_{\ell})$ by almost two orders of magnitude.}
\label{fig:mcplot}
\end{center}
\end{figure}

In summary the average logical error rate is grossly underestimated unless an unreasonably large number of outcomes are sampled. We will resort to an importance sampling technique proposed in \cite{IP17}, to improve our estimate of the average logical error rate. Previously, similar techniques have also been discussed for Pauli noise processes in \cite{BV13,TLGD17}. Instead of choosing to sample the syndrome probability distribution, we sample an alternate distribution $Q(s)$, which we will simply refer to as the \emph{importance distribution}. The corresponding sampling methods with $\Pr(s)$ and $Q(s)$ will be referred to as \emph{direct sampling} and \emph{importance sampling} respectively.

The expression for the average logical error rate estimated by the importance sampler takes a form:
\begin{gather}
r(\hat{\cE}_{\ell}) = \sum_{\hat{s}} r(\cE^{\hat{s}}_{\ell}) \dfrac{\Pr(s)}{Q(s)}\ths , \label{eq:avgchan_is}
\end{gather}
where $\hat{s}$ is a random syndrome outcome drawn from the importance distribution $Q(s)$. The average estimated by importance sampling coincides with $r(\ol{\cE}_{\ell})$ which is estimated by the direct sampling technique. The crucial difference between the two sampling techniques is that the variance of the estimated average can be significantly lowered by an appropriate choice for the importance distribution $Q(s)$, which in our case, takes the form
\begin{gather}
Q(s) = \dfrac{P(s)^{1/k}}{Z} \ths , \label{eq:impdist}
\end{gather}
where $Z$ is a normalization factor
\begin{gather}
Z = \sum_{s}P(s)^{1/k} \ths , \label{eq:impdist_norm}
\end{gather}
and $k \in (0,1]$ is chosen such that the total probability of non-trivial syndrome outcomes, $s \neq 00\ldots 0$, is above a fixed threshold $\lambda_{0}$, i.e.,
\begin{gather}
\sum_{s \neq 00\ldots 0} \dfrac{\Pr(s)^{1/k}}{Z} \geq \lambda_{0} \ths . \label{eq:impdist_expo}
\end{gather}
Figure \ref{fig:mcplot} shows that our heuristic for the importance distribution provides a rapid convergence to $r(\ol{\cE}_{\ell})$, when compared to the direct sampling method. Note that the noise processes in these figures are the same as those used to compare the predictive powers of physical error metrics in figures \ref{fig:pred} and \ref{fig:pred_coherent}. Hence, the employment of importance sampling is key to an honest comparison of the predictive powers of the physical error metrics.
\end{widetext}
\end{appendix}

\bibliographystyle{unsrt}
\bibliography{refs}

\end{document}